\documentclass[12pt,showpacs,preprintnumbers,amsmath,aps,amssymb]{revtex4}
\usepackage{graphicx}% Include figure files
\usepackage{dcolumn}% Align table columns on decimal point
\usepackage{bm}% bold math
\usepackage{color}
\usepackage{xcolor}
\usepackage{multirow}
\usepackage{amsmath,amsfonts}

\begin{document}

\baselineskip 18pt

\title{Non commutative classical and  Quantum fractionary Cosmology: Anisotropic Bianchi Type I case.}
\author{J. Socorro$^{1}$ }
\email{socorro@fisica.ugto.mx}
\author{J. Juan Rosales $^2$}
\email{rosales@ugto.mx}
\author{Leonel Toledo-Sesma$^3$}
\email{ltoledos@ipn.mx} \affiliation{$^{1}$ \quad Department of
Physics,  Division of Science and Engineering, University of
Guanajuato, Campus Le\'{o}n,
 Le\'{o}n 37150,  Mexico\\
$^{2}$ \quad Department of Electrical Engineering, Engineering
Division Campus Irapuato-Salamanca, University of Guanajuato,
Salamanca 36885, Mexico;
 rosales@ugto.mx\\
$^{3}$ \quad National Polytechnic Institute, UPIIH, Carretera
Pachuca---Actopan Kilometer 1+500, San\mbox{ Agust\'in Tlaxiaca
42162},  Hidalgo, M\'exico; ltoledos@ipn.mx}

\begin{abstract}
In this work, we will explore the effects of non-commutativity in
fractional classical and quantum schemes using the anisotropicc
Bianchi Type I cosmological model coupled to a scalar field in the
K-essence formalism. We introduce non-commutative variables
considering that all minisuperspace variables $q^i_{nc}$ do not
commute, so the symplectic structure was modified, resulting in some
changes with respect to the traditional formalism. In the quantum
regime, the probability density presents a new structure in the
scalar field corresponding to the value of the non-commutative
parameter.


\end{abstract}

\maketitle

\section{Introduction}

Many works investigate the Fractional Calculus (FC) and its
applications \cite{r-s}, being a powerful mathematical tool for
describing complex processes, such as the tautochrone problem
\cite{Abel}, models based on memory mechanism \cite{Caputo},
anomalous diffusion \cite{Wyss}, linear capacitor theory
\cite{Westerlund}, non-local description of quantum mechanics
\cite{Hermann},
 processing of medical images \cite{Jorge,Leo},
among others. Recently, CF has been applied to the general theory of
relativity
\cite{co1,co2,co3,co4,co5,co6,co7,co8,co9,co10,co11,co12,paulo3,paulo4,paulo5},
quantum mechanics and quantum cosmology \cite{paulo1,paulo2}; with
interesting results in cosmology, such as that the age of the
Universe is greater than or of the order of 13.8196 Gyr,
\cite{Jalalzadeh2023}. More recently, a new fractal cosmological
model $\Lambda$CDM has been created, where it is shown that the
fractal spatial dimension is two and the age of the Universe is
13.91 Gyr, \cite{Jalalzadeh2024}. The study and applications of
fractional calculus  to cosmology is a new line of research and we
have recently worked along this line in the theory of K-essence
where we find several relevant indicators for its study. When this
theory is analyzed as a perfect fluid and in particular the
barotropic parameter is constant, it is generally demonstrated that
this theory is equivalent to general relativity coupled to ordinary
matter with a barotropic equation of state \cite{sasaki2010}, which
has been verified in particular with the standard FRW model
\cite{universe} and an anisotropic model, the Bianchi type I. The
second indicator relevant to our case is that this is the only
theory that when quantized under the ADM formalism, a fractional
Wheeler-DeWitt (WDW) equation, in the scalar field component, is
obtained naturally at different stages of the universe
\cite{universe}. We have recently worked on non-commutativity (NC)
in the space phase in this formalism \cite{universe2024-nc}. It is
well known that there are various ways to introduce
non-commutativity in the phase space and that they produce different
dynamical systems from the same Lagrangian \cite{Abreu-2006}, as can
be shown for example in reference \cite{De-andrade} and references
cited therein. Therefore, distinct choices for the NC algebra among
the brackets render distinct dynamic systems. We will use
non-commutativity in the coordinate space, leaving the application
of moments space for the future, \cite{Guzman-2011},
 where other quantities such as angular momentum appear between coordinates and momenta \cite{sabido2018,sabido2024}.
In this work,  the known mathematical structure on non-commutativity
undergoes relevant changes since a prefactor appears that was
previously absorbed by the choice of the gauge. At the classical
level this prefactor does not cause a problem, since several of the
moments turn out to be constant, but at the quantum level we do have
to make an adequate separation to have something reasonable
physically, and will depend strongly on how it is used  to reproduce
plausible results that allow us to recover the  commutative quantum
world, however, the combination of fractional differential equations
and non-commutativity in anisotropic models seems to be not a good
combination, since when solving the resulting quantum equations,
this combination causes the domain of definition of the
non-commutative parameter to change to another domain of definition
that does not allow us to viably recover the commutative quantum
world. In this way,  we discarded the other choices of the prefactor
given that due to the way in which the non-commutative parameter is
worked, it seems that we are facing an ambiguity or separation of
definition space.

Let us consider the Misner parameterization \cite{misner} in the
line element of Bianchi-type I cosmological models
\begin{eqnarray}
ds^2&=& -N^2 dt^2 + A^2 dx^2 + B^2 dy^2 + C^2 dz^2,
\label{traditional}\\
&=& -N^2 dt^2 + e^{2\Omega + 2\beta_++2\sqrt{3}\beta_-}dx^2 +
e^{2\Omega + 2\beta_+-2\sqrt{3}\beta_-}dy^2 + e^{2\Omega -
4\beta_+}dz^2, \label{misner-p}
\end{eqnarray}
where $N(t)$ is the lapse function, $\Omega(t)$ is a scalar
function. We use the Misner's parametrization, then the radii are
given by

$$A=e^{\Omega + \beta_++\sqrt{3}\beta_-}, \qquad =e^{\Omega + \beta_+-\sqrt{3}\beta_-}, \qquad C=e^{\Omega-2\beta_+}.$$
and the K-essence Lagrangian density of the form \cite{1,roland,chiba,bose,arroja,tejeiro}

\begin{equation}
S=\int d^4x \, \sqrt{-g}\,\left[ f(\phi) \, {\cal G}(X)\right],
\end{equation}
where the canonical kinetic energy is given by
$\mathcal{G}(X)=X=-\frac{1}{2}\nabla_\mu \phi \nabla^\mu \phi$, $R$
is the scalar of curvature, $f(\phi)$ is an arbitrary function of
the scalar field $\phi$. In this work, we choose $f(\phi)=1$, and
$g$ is the determinant of the metric tensor. The K-essence model has
been proposed as an inflationary model, also as a model for dark
energy along with explorations to unify energy and dark matter
\cite{roland,bilic,bento}. Another motivation to consider this type
of Lagrangian originates from string theory \cite{string}. For more
details on K-essence applied to dark energy, see \cite{copeland} and
references therein.

The simplest  K-essence Lagrangian density is

\begin{equation}
\mathcal{L}_{geo}=\sqrt{-g}\left( \frac{R}{2}+   {\cal G}(X)\right),
\label{lagrangian}
\end{equation}
hence, the field equations are given by

\begin{subequations}
\begin{eqnarray}
G_{\alpha \beta}+  \left[{\cal G}_X \phi_{,\alpha}\phi_{,\beta}
 +  {\cal G} g_{\alpha \beta}  \right]
&=& T_{\alpha \beta}, \label{Efe} \\
 \left[{\cal G}_X\phi^{,\beta}_{\, ;\beta} + {\cal
G}_{XX}X_{;\beta}\phi^{,\beta} \right]&=&0, \label{Sfe}
\end{eqnarray}
\end{subequations}
where, we have assumed the units with $8\pi G=1$, the semicolon
means a covariant derivative and the subscript $X$ denotes
differentiation with respect to $X$.

The same set of equations \eqref{Efe} and \eqref{Sfe} are obtained
if we consider the scalar field $X(\phi)$ as part of the matter
content,  \textit{i.e.} $\mathcal{L}_{X,\phi} = \mathcal{G}(X)$,
with the corresponding energy-momentum tensor

\begin{equation}
\mathcal{T}_{\alpha \beta}= \left[\mathcal{G}_X
\phi_{,\alpha}\phi_{,\beta}
 +  \mathcal{G}(X)g_{\alpha \beta}  \right], \label{ener-mom}
\end{equation}
also, considering the energy-momentum tensor of a barotropic perfect
fluid,
\begin{equation}
T_{\alpha\beta}=(\rho +P)u_\alpha u_\beta + P g_{\alpha \beta},
\label{tensor-matter}
\end{equation}
with $u_\alpha$ being the four-velocity satisfying the relation
$u_\mu u^\mu=-1$, $\rho$ the energy density and $P$ the pressure
of the fluid. For simplicity, we consider a comoving perfect fluid.
The pressure and energy density corresponding to the energy
momentum tensor of the field $X$, are

\begin{equation}
P(X)=  {\cal G}, \qquad \rho(X)= \left[2X{\cal G}_X-{\cal G}
\right], \label{pX}
\end{equation}
thus, the barotropic parameter $\omega_X= \frac{P(X)}{\rho(X)} $ for
the equivalent fluid  is given by

\begin{equation}
\omega_X=\frac{\mathcal{G}}{2X\mathcal{G}_X - \mathcal{G}}.
\label{ma}
\end{equation}
The case of a constant barotropic parameter $\omega_X$, (with the
exception $\omega_X=0$) can be obtained by the solution to equation
(\ref{ma}),
 $\mathcal{G}$ function
\begin{equation}
\mathcal{G}=X^{\frac{1+\omega_X}{2\omega_X}},
\end{equation}
choosing the barotropic parameter as
$$\omega_X=\frac{2\kappa-1}{2\kappa+1}, \qquad \to \qquad \mathcal{
G}=X^\alpha, $$ where the $\alpha$ parameter
\begin{equation}
\alpha =\frac{2\kappa}{2\kappa-1},
\label{fc-parameter}
\end{equation}
will be relevant in our study. Thus, we can write the barotropic
parameter in terms of $\alpha$ as $\omega_X=\frac{1}{2\alpha -1}$,
being $\kappa=\frac{\alpha}{2(\alpha-1)}$. Taking into account the
above, we can write the states in the evolution of our universe in
the table \ref{tab1}, where we have introduced the $\beta$ parameter
that has to do with the fractional order of the Wheeler-DeWitt
equation in the scalar field.

\begin{table}[!htbp]
    \centering
  \begin{tabular}{ | c | c | c | c | c | c|}
  \hline \hline
  States & $\kappa$ & $\alpha$&$\beta=\frac{2\alpha}{2\alpha-1}$ & $\omega_{X}=\frac{1}{2\alpha-1}$ & Kinetic energy $(\mathcal{G}(X))$ \\
  \hline \hline
     Stiff matter & $\to \infty$ & 1 & 2 &1 & $X$ \\
     \hline
     Radiation & 1 & 2 & $\frac{4}{3}$ & $\frac{1}{3}$ & $X^{2}$ \\
     \hline
     Dust-like & $\to \frac{1}{2}$ & $\to\infty$& 1 & $\to 0$ & $X^{m}, \qquad m\to\infty$ \\
     \hline
     Inflation & 0 & 0 & 0 & -1 & $1, \qquad \Lambda = cte.$ \\
     \hline
     Inflation-like & $\frac{1}{4}$ &  -1&$\frac{2}{3}$& $-\frac{1}{3}$ & $\frac{1}{X}$ \\
     %\hline
                         & $\frac{1}{10}$ &$-\frac{1}{4}$&$\frac{1}{3} $& $-\frac{2}{3}$ & $\frac{1}{\sqrt[4]{X}}$ \\
    \hline
  \end{tabular}
  \caption{States of universe's evolution.}
  \label{tab1}
\end{table}

\section{Lagrange and Hamilton formalism}

Introducing the line element \eqref{misner-p} into the Lagrangian
\eqref{lagrangian}, we get

\begin{equation}
\mathcal{L}_I= e^{3\Omega}\left[6\frac{\dot \Omega^2}{N} -6\frac{{\dot
\beta_+}^2}{N}-6\frac{{\dot \beta_-}^2}{N}
-\,\left(\frac{1}{2}\right)^\alpha \left( \dot \phi
\right)^{2\alpha}\,N^{-2\alpha+1} \right]. \label{lagra-vi}
\end{equation}
Using the standard definition of the momenta
$\Pi_{q^{\mu}}=\frac{\partial\mathcal{L}}{\partial\dot{q}^{\mu}}$,
where $q^{\mu}$ are the coordinate fields
$q^{\mu}=(\Omega,\beta_+,\beta_-,\phi)$, we obtain the momenta
associated to each field;

\begin{align}
\Pi_{\Omega} & =12e^{3\Omega}\frac{\dot \Omega}{N},  &   \dot{\Omega} & = Ne^{-3\Omega}\frac{\Pi_\Omega}{12}, \nonumber\\
\Pi_{+} & = -12e^{3\Omega}\frac{\dot \beta_+}{N}, &  \dot{\beta_+} & = -Ne^{-3\Omega}\frac{ \Pi_+}{12}, \nonumber\\
\Pi_{-} & = -12e^{3\Omega}\frac{\dot \beta_-}{N}, &  \dot{\beta_-} & = -Ne^{-3\Omega}\frac{ \Pi_-}{12}, \\
\Pi_\phi & =
-\,\left(\frac{1}{2}\right)^\alpha\frac{2\alpha}{N^{2\alpha-1}}\,e^{3\Omega}{\dot
\phi}^{2\alpha -1}, & \dot{\phi} & = -N\left[\frac{2^{\alpha-1}}{\alpha
e^{3\Omega}}\Pi_\phi\right]^{\frac{1}{2\alpha -1}},\nonumber
\label{ph}
\end{align}
and introducing them into the Lagrangian density, we obtain the canonical Lagrangian as $\mathcal{L}_{canonical} =\Pi_{q^\mu} \dot
q^\mu - N\mathcal{H}$. When we perform the variation of this canonical Lagrangian with respect to $N$, $\frac{\delta\mathcal{L}_{canonical}}{\delta N} =0$,
 we obtain the constraint $\mathcal{H}\approx 0$.

 Then, we obtain the corresponding Hamiltonian density
\begin{equation}
\mathcal{H}_I=
N\frac{e^{-\frac{3}{2\alpha-1}\Omega}}{24}\left\{e^{-\frac{6(\alpha-1)}{2\alpha-1}\Omega}
\left[\Pi_\Omega^2- \Pi_+^2-\Pi_-^2
 \right]
 -\frac{12(2\alpha-1)}{\alpha }\,\left(
\frac{2^{\alpha-1}}{\alpha
}\right)^{\frac{1}{2\alpha-1}}\Pi_\phi^{\frac{2\alpha}{2\alpha-1}}\right\}.
\label{ham}
\end{equation}

Next we take the gauge $N=24e^{\frac{3}{2\alpha 1}\Omega}$ and we
will continue the analysis with the reduced Hamiltonian

\begin{equation}
\mathcal{H}=e^{-\frac{6(\alpha-1)}{2\alpha-1}\Omega} \left[\Pi_\Omega^2- \Pi_+^2-\Pi_-^2  \right]
 -\frac{12(2\alpha-1)}{\alpha }\,\left( \frac{2^{\alpha-1}}{\alpha
}\right)^{\frac{1}{2\alpha-1}}\Pi_\phi^{\frac{2\alpha}{2\alpha-1}}.
\label{reducido}
\end{equation}
The corresponding Hamilton equations become

\begin{eqnarray}
\dot \Omega &=& 2\Pi_\Omega e^{-\frac{6(\alpha-1)}{2\alpha-1}
\Omega},\label{domega}\\
\dot \beta_\pm &=& -2\Pi_\pm e^{-\frac{6(\alpha-1)}{2\alpha-1}
\Omega},\label{dmm}\\
\dot \phi &=& -24 \left( \frac{2^{\alpha-1}}{\alpha
}\right)^{\frac{1}{2\alpha-1}}
\Pi_\phi^{\frac{1}{2\alpha-1}}, \label{dphi}\\
\dot \Pi_\Omega &=&\frac{6(\alpha-1)}{2\alpha-1}\,
e^{-\frac{6(\alpha-1)}{2\alpha-1} \Omega} \left[\Pi_\Omega^2
-\Pi_+^2 -\Pi_-^2  \right]
\label {dpio}\\
\dot \Pi_+ &=& 0,\qquad \Pi_+=p_+=constant, \label {dpi+}\\
\dot \Pi_- &=& 0,\qquad \Pi_-=p_-=constant, \label {dpi-}\\
\dot \Pi_\phi &=& 0, \qquad \Pi_\phi= p_\phi=constant,
\label{dpiphi}
\end{eqnarray}
substituting \eqref{dpiphi} into \eqref{dphi} we have the time
solution for the scalar field as

\begin{equation}
\phi(t)=\phi_i -24 \left( \frac{2^{\alpha-1}}{\alpha
}\right)^{\frac{1}{2\alpha-1}}
p_\phi^{\frac{1}{2\alpha-1}}\left(t-t_{\alpha_i}\right)
\end{equation}
where $\phi_i$ and  $t_{\alpha_i}$ are the initial value for the scalar field and  the initial time in the $\alpha$ scenario, respectively.
\\

On the other hand, using the Hamiltonian constraint
(\ref{reducido}), we found

\begin{equation}
e^{-\frac{6(\alpha-1)}{2\alpha-1}\Omega} \left[\Pi_\Omega^2-
\Pi_+^2-\Pi_-^2
 \right]
 =\frac{12(2\alpha-1)}{\alpha }\,\left(
\frac{2^{\alpha-1}}{\alpha
}\right)^{\frac{1}{2\alpha-1}}p_\phi^{\frac{2\alpha}{2\alpha-1}}
=constant,
\end{equation}
and substituting into equations \eqref{dpio} we have the time
solution

\begin{equation}
\Pi_\Omega(t)=p_0+\frac{72(\alpha-1)}{\alpha}\left(
\frac{2^{\alpha-1}}{\alpha
}\right)^{\frac{1}{2\alpha-1}}p_\phi^{\frac{2\alpha}{2\alpha-1}}
\left(t-t_{\alpha_0} \right),\label{solucion-pi-o}
\end{equation}
where $p_0$ is an integration constant.

The equation \eqref{domega} can be rewritten as
\begin{equation}
\left(
\frac{2\alpha-1}{6(\alpha-1)}e^{\frac{6(\alpha-1)}{2\alpha-1}\Omega}\right)^\bullet
=2\Pi_\Omega,
\end{equation}
substituting the solution for $\Pi_\Omega(t)$, we get

\begin{equation}
e^{\frac{6(\alpha-1)}{2\alpha-1}\Omega}=a_0+\frac{12(\alpha-1)}{2\alpha-1}p_0(t-t_{\alpha_0})+\frac{432(\alpha-1)^2}{\alpha(2\alpha-1)}\left(
\frac{2^{\alpha-1}}{\alpha
}\right)^{\frac{1}{2\alpha-1}}p_\phi^{\frac{2\alpha}{2\alpha-1}}
\left(t-t_{\alpha_0} \right)^2.
\end{equation}
Hence, the $\Omega(t)$ becomes

\begin{equation}
\Omega(t)=\frac{2\alpha-1}{6(\alpha-1)}\, Ln
\left[a_0+\frac{12(\alpha-1)}{2\alpha-1}p_0(t-t_{\alpha_0})+\frac{432(\alpha-1)^2}{\alpha(2\alpha-1)}\left(
\frac{2^{\alpha-1}}{\alpha
}\right)^{\frac{1}{2\alpha-1}}p_\phi^{\frac{2\alpha}{2\alpha-1}}
\left(t-t_{\alpha_0} \right)^2 \right]
\end{equation}
and the corresponding solutions for the anisotropic parameter $\beta_\pm$ are

\begin{equation}
\beta_\pm(t)=\beta_{{\pm}_{0}}-2 p_\pm\,\int
\frac{dz}{a_0+b_\alpha\,z+c_\alpha z^2}, \qquad z=t-t_{\alpha_0}.
\end{equation}
with $b_\alpha=\frac{12(\alpha-1)}{2\alpha-1}p_0,
c_\alpha=\frac{432(\alpha-1)^2}{\alpha(2\alpha-1)}\left(
\frac{2^{\alpha-1}}{\alpha}\right)^{\frac{1}{2\alpha-1}}p_\phi^{\frac{2\alpha}{2\alpha-1}}$,
and ${\beta_0}_\pm$ are integration constants, so, the solution
becomes

\begin{equation*}
\int \frac{dz}{a_0+b_\alpha\,z+c_\alpha z^2}=\frac{1}{\sqrt{b_\alpha^2-4a_0c_\alpha}}\,
\ln{ \left[\frac{2c_\alpha(t-t_{\alpha_0})+b_\alpha-
\sqrt{b_\alpha^2-4a_0c_\alpha}}{2c_\alpha(t-t_{\alpha_0})+b_\alpha+
\sqrt{b_\alpha^2-4a_0c_\alpha}} \right]},
\end{equation*}
with the condition over the constants $b_\alpha
>2\sqrt{a_0 c_\alpha}$.

As we have mentioned, the Hamiltonian (\ref{reducido}) has parts
that are fractional derivatives and parts that are integers. The
case of integer derivatives must be resolved in a particular way.
%%%%%%%%%%
\subsection{Stiff matter scenario, $\alpha=1$, $\beta=2$.}

When we analyze the stiff matter scenario for the universe follows
from the Hamiltonian constriction (\ref{reducido})

\begin{equation}
\mathcal{H} = \Pi_\Omega^2 - \Pi_+^2 - \Pi_-^2 -12\Pi_\phi^2,
\label{stiff-matter}
\end{equation}
we have the following Hamilton equations

\begin{eqnarray}
\dot \Omega &=& 2\Pi_\Omega ,\label{domega1}\\
\dot \beta_\pm &=&-2\Pi_{\pm}, \label{1}\\
\dot \phi &=& -24
\Pi_\phi, \label{dphi1}\\
\dot \Pi_\Omega &=& 0, \qquad \Pi_\Omega=p_0=constant, \label {dpio1}\\
\dot \Pi_\pm&=&0, \qquad \Pi_\pm=p_\pm=constants, \label {dpi+1}\\
\dot \Pi_\phi &=& 0, \quad \Pi_\phi=p_\phi = constant,
\label{dpiphi1}
\end{eqnarray}
Then, the $\Omega$ solution becomes

\begin{equation}
\Omega(t)=\Omega_0 +2p_0(t-t_s).\label{sca}
\end{equation}
and the anisotropic variables have the solutions
$\beta_\pm(t)=\beta_{\pm_0}-2p_\pm(t-t_s)$. The volume of the
universe for this scenario  is, $V(t)=e^{3\Omega(t)}$, given by

\begin{equation}
V(t)=V_0\,e^{6p_0(t-t_s)}.
\end{equation}
And the solution for the scalar field $\phi$ is

\begin{equation}
\phi(t) =\phi_s-24 p_\phi (t-t_s).
\end{equation}
where $\phi_s$ is the initial value in the stiff matter scenario.
%%%%%%%%%%
\subsection{Dust scenario, $\alpha \to \infty$, $\beta=1$.}
For this scenario, the Hamiltonian constraint (\ref{reducido}) is
\begin{equation}
\mathcal{H}=e^{-3\Omega} \left[\Pi_\Omega^2 - \Pi_+^2  - \Pi_-^2\right] -12\sqrt{2}\Pi_\phi,
\label{dust}
\end{equation}
with the Hamilton equations

\begin{eqnarray}
\dot \Omega&=& 2e^{-3\Omega}\Pi_\Omega, \label{dust-omega}\\
\dot \beta_\pm &=& -2e^{-3\Omega}\Pi_\pm, \label{dust-beta}\\
\dot \phi &=& -12\sqrt{2}, \qquad \phi(t)=\phi_d -12\sqrt{2}(t-t_d),
\label{dust-phi}
\\
\dot \Pi_\Omega &=& 3e^{-3\Omega}\left[\Pi_\Omega^2- \Pi_+^2
 - \Pi_-^2\right], \label{dust-pi-omega}\\
 \dot \Pi_\pm &=& 0,\qquad \Pi_\pm =p_\pm =constant, \label{dust-pi-beta}\\
 \dot \Pi_\phi &=& 0, \qquad \Pi_\phi=p_\phi=constant. \label{dust-pi-phi}
 \end{eqnarray}
 where the constant $\phi_d$ must be very huge in such a way that this scalar field survives actually.

Now, using the Hamiltonian constraint, we found that $\Pi_\Omega(t)=p_d+36\sqrt{2}(t-t_d)$,
and substituting into \eqref{dust-omega}, we have that the volume of this universe becomes

\begin{equation}
V(t)=e^{3\Omega}=a_0+6p_d(t-t_d)+108\sqrt{2}p_\phi(t-t_d)^2,
\label{vol-dust}
\end{equation}
with $a_0$ an integration constant, and the other solutions are

\begin{equation}
\beta_\pm={\beta_0}_{\pm}-2p_\pm \int \frac{dz}{a_0+b_0 z+ c_0z^2},
\qquad z=t-t_d.
\label{parameter-ani}
\end{equation}
where $b_0=6p_d$ and $c_0=108\sqrt{2}p_\phi$, and the integral

\begin{equation*}
\int\frac{dz}{a_0+b_0 z+
c_0z^2}=\frac{1}{\sqrt{b_0^2-4a_0c_0}}
\ln{\left[\frac{2c_0(t-t_d)+b_0-\sqrt{b_0^2-4a_0c_0}}{2c_0(t-t_d)+b_0+\sqrt{b_0^2-4a_0c_0}}
 \right]}
\end{equation*}
with the condition over the constants $36p_d^2-432a_0\sqrt{2}p_\phi>0$, and $p_d>2\sqrt{3\sqrt{2}a_0p_\phi}$.

\section{Non-commutative fractional classical exact solution}\label{noncommutative}

Rewriting the classical Hamiltonian (\ref{reducido}) in terms of the
fractional parameter $\beta=\frac{2\alpha}{2\alpha-1}$, we find the
commutative equation of motion in the classical phase space
variables $ q^\mu=(\Omega,\beta_+,\beta_-,\phi)$, with the Poisson
algebra:
\begin{equation}
 \left\{ q^\mu, q^\nu  \right\}=0 \qquad \left\{ \Pi_{q^\mu}, \Pi_{q^\nu} \right\}=0, \qquad \left\{q^\mu,\Pi_{q^\nu}
\right\}=\delta^\mu_\nu,
\label{cbracket}
\end{equation}

\begin{equation}
\mathcal{H}= e^{-3(2-\beta)\Omega}\left[\Pi_\Omega^2- \Pi_+^2-\Pi_-^2\right] - \frac{24}{\beta}\left(\frac{2^{\alpha-1}}{\alpha\, f(\phi)}
\right)^{\frac{1}{2\alpha-1}}\,\Pi_\phi^{\beta},
\label{hami-beta}
\end{equation}

Now, the natural extension is to consider the non-commutative version of our model, with the idea of non-commutative between all
variables $(\Omega_{nc},{\beta_\pm}_{nc},\phi_{nc})$, so we apply a deformation of the Poisson algebra. For this, we start with
the usual Hamiltonian \eqref{hami-beta}, but the symplectic structure is modified as follows

\begin{align}
\left\{\Pi_{\Omega},\Pi_\phi\right\}_\star & =0, & \left\{q^\mu,\Pi_{q^\mu} \right\}_\star & = 1, & \left\{\Omega,\beta_+\right\}_\star & = \theta_1, & \left\{\Omega,\beta_-\right\}_\star & = \theta_2, \nonumber\\
\left\{\Omega,\phi\right\}_\star & = \theta_3, & \left\{\beta_+,\beta_-\right\}_\star & = \theta_4, & \left\{\beta_+,\phi\right\}_\star & = \theta_5, & \left\{\beta_-,\phi\right\}_\star & = \theta_6,
\label{ncbracket}
\end{align}
where the $\star$ is the Moyal product \cite{Szabo2} and  the
resulting Hamiltonian density is

\begin{equation}
\mathcal{H}_{nc}= e^{-3(2-\beta)\Omega_{nc}}\left[\Pi_\Omega^2 - \Pi_+^2 - \Pi_-^2\right] - \frac{24}{\beta}\left(\frac{2^{\alpha-1}}{\alpha\, }
\right)^{\frac{1}{2\alpha-1}}\,\Pi_\phi^{\beta},
\label{hami-nc}
\end{equation}

When we used the sifted variables (Bopp shift approach) but with the original (commutative) symplectic structure

\begin{equation}
\dot q^\mu_{nc} = \left\{q^\mu_{nc}, \mathcal{H}_{nc} \right\}, \qquad \dot {\Pi_\mu}_{nc} = \left\{{\Pi_\mu}_{nc}, \mathcal{H}_{nc}\right\},
\end{equation}
the commutation relations \eqref{ncbracket} can be implemented in
terms of the commutation coordinates and it results in a
modification of the potential-like terms in the Hamiltonian density.
In other words, this same result can be found employing the Bob
Shift between the commutative coordinates
$q^\mu=(\Omega,\beta_+,\beta_-,\phi)$ and the non-commutative
coordinates $q^\mu_{nc}=(\Omega_{nc},
\beta_+{_{nc}},\beta_-{_{nc}},\phi_{n})$ by the relation

\begin{equation}
q^\mu= q^\mu_{nc}+\frac{1}{2}\Theta^{\mu \nu} \Pi_\nu,
\label{relation}
\end{equation}
where we choose the matrix $\Theta^{\mu \nu}$ as
\begin{equation}
\Theta^{\mu \nu} = \left( \begin{tabular}{cccc} 0 & $\theta_1$ &
$\theta_2$ & $\theta_3$ \\
$-\theta_1$ & 0 &
$\theta_4$ & $\theta_5$ \\
$-\theta_2$ & $-\theta_4$ & 0 & $\theta_6$\\
 $-\theta_3$ &
$-\theta_5$ & $-\theta_6$ & 0
\end{tabular}
 \right)
 \end{equation}
 we have, then

\begin{eqnarray}
\Omega &=& \Omega_{nc}+\frac{\theta_1}{2}\Pi_+ +
\frac{\theta_2}{2}\Pi_- + \frac{\theta_3}{2}\Pi_\phi,\label{omega}\\
\beta_+ &=& \beta_+{_{nc}} -\frac{\theta_1}{2}\Pi_\Omega
+\frac{\theta_4}{2}\Pi_- +\frac{\theta_5}{2}\Pi_\phi,
\label{beta+}\\
\beta_- &=& \beta_-{_{nc}} -\frac{\theta_2}{2}\Pi_\Omega
-\frac{\theta_4}{2}\Pi_+ +\frac{\theta_6}{2}\Pi_\phi,
\label{beta-}\\
\phi &=& \phi_{nc}
-\frac{\theta_3}{2}\Pi_\Omega-\frac{\theta_5}{2}\Pi_+-\frac{\theta_6}{2}\Pi_-.
\label{phi}
\end{eqnarray}
Therefore

\begin{eqnarray}
\Omega_{nc} &=& \Omega -\frac{\theta_1}{2}\Pi_+ -
\frac{\theta_2}{2}\Pi_- - \frac{\theta_3}{2}\Pi_\phi,\label{omega-nc}\\
\beta_+{_{nc}} &=& \beta_+ +\frac{\theta_1}{2}\Pi_\Omega
-\frac{\theta_4}{2}\Pi_- -\frac{\theta_5}{2}\Pi_\phi,
\label{beta+nc}\\
\beta_-{_{nc}} &=& \beta_- +\frac{\theta_2}{2}\Pi_\Omega
+\frac{\theta_4}{2}\Pi_+ -\frac{\theta_6}{2}\Pi_\phi,
\label{beta-0}\\
\phi_{nc} &=& \phi
+\frac{\theta_3}{2}\Pi_\Omega+\frac{\theta_5}{2}\Pi_++\frac{\theta_6}{2}\Pi_-.
\label{phi00}
\end{eqnarray}

Therefore, our Hamiltonian density has the form
\begin{equation}
\mathcal{H}_{nc}= e^{-3(2-\beta)\left[\Omega - \frac{\theta_1}{2}\Pi_+
-\frac{\theta_2}{2}\Pi_- -
\frac{\theta_3}{2}\Pi_\phi\right]}\left[\Pi_\Omega^2-
\Pi_+^2-\Pi_-^2
 \right] -
\frac{24}{\beta}\left(\frac{2^{\alpha-1}}{\alpha }
\right)^{\frac{1}{2\alpha-1}}\,\Pi_\phi^{\beta}. \label{hami-nc-n}
\end{equation}
The Hamilton equations are

\begin{eqnarray}
\dot \Omega &=& 2\Pi_\Omega
e^{-3(2-\beta)\left[\Omega-\frac{\theta_1}{2}\Pi_+
-\frac{\theta_2}{2}\Pi_- -\frac{\theta_3}{2}\Pi_\phi\right]},
\label{o-nc}\\
\dot \beta_+ &=&
-2\Pi_+\,e^{-3(2-\beta)\left[\Omega-\frac{\theta_1}{2}\Pi_+
-\frac{\theta_2}{2}\Pi_-
-\frac{\theta_3}{2}\Pi_\phi\right]}\nonumber\\
&& \qquad +\theta_1\,\frac{3(2-\beta)}{2}e^{-3(2-\beta)\left[\Omega
- \frac{\theta_1}{2}\Pi_+ -\frac{\theta_2}{2}\Pi_- -
\frac{\theta_3}{2}\Pi_\phi\right]}\left[\Pi_\Omega^2-
\Pi_+^2-\Pi_-^2
 \right], \label{+-nc}\\
\dot \beta_- &=&
-2\Pi_-\,e^{-3(2-\beta)\left[\Omega-\frac{\theta_1}{2}\Pi_+
-\frac{\theta_2}{2}\Pi_-
-\frac{\theta_3}{2}\Pi_\phi\right]}\nonumber\\
&&\quad +\theta_2\,\frac{3(2-\beta)}{2}e^{-3(2-\beta)\left[\Omega -
\frac{\theta_1}{2}\Pi_+ -\frac{\theta_2}{2}\Pi_- -
\frac{\theta_3}{2}\Pi_\phi\right]}\left[\Pi_\Omega^2-
\Pi_+^2-\Pi_-^2
 \right], \label{+-nc00}\\
 \dot \phi &=& -24 \frac{24}{\beta}\left(\frac{2^{\alpha-1}}{\alpha }
\right)^{\frac{1}{2\alpha-1}}\,\Pi_\phi^{\beta-1}\nonumber\\
&&\quad +\theta_3\,\frac{3(2-\beta)}{2}e^{-3(2-\beta)\left[\Omega -
\frac{\theta_1}{2}\Pi_+ -\frac{\theta_2}{2}\Pi_- -
\frac{\theta_3}{2}\Pi_\phi\right]}\left[\Pi_\Omega^2-
\Pi_+^2-\Pi_-^2 \right], \label{phi-nc}\\
\dot \Pi_\Omega &=&
3(2-\beta)e^{-3(2-\beta)\left[\Omega-\frac{\theta_1}{2}\Pi_+
-\frac{\theta_2}{2}\Pi_-
-\frac{\theta_3}{2}\Pi_\phi\right]}\left[\Pi_\Omega^2-
\Pi_+^2-\Pi_-^2
 \right],
\label{piomega-nc}\\
\dot \Pi_\pm &=& 0, \qquad \Pi_\pm=p_\pm=constants,
\label{pipm-nc}\\
\dot \Pi_\phi &=& 0, \qquad \Pi_\phi=p_\phi=constant.
\label{piphi-nc}
\end{eqnarray}
Similarly, using the Hamiltonian constraint and partial results in
the Hamilton equations, we find, that

\begin{equation*}
e^{-3(2-\beta)\left[\Omega-\frac{\theta_1}{2}\Pi_+
-\frac{\theta_2}{2}\Pi_-
-\frac{\theta_3}{2}\Pi_\phi\right]}\left[\Pi_\Omega^2-
\Pi_+^2-\Pi_-^2
 \right]=\frac{24}{\beta}\left(\frac{2^{\alpha-1}}{\alpha }
\right)^{\frac{1}{2\alpha-1}}\,p_\phi^{\beta}=constant,
\end{equation*}
then

\begin{equation}
\dot
\Pi_\Omega=\frac{72(2-\beta)}{\beta}\left(\frac{2^{\alpha-1}}{\alpha
} \right)^{\frac{1}{2\alpha-1}}\,p_\phi^{\beta}, \qquad
\Pi_\Omega=p_0
+\frac{72(2-\beta)}{\beta}\left(\frac{2^{\alpha-1}}{\alpha }
\right)^{\frac{1}{2\alpha-1}}\,p_\phi^{\beta}\left(t-t_{\alpha_i}
\right).
\end{equation}
the $\Omega$ equation can be rewritten as

\begin{equation*}
\dot \Omega=2\Pi_\Omega e^{-3(2-\beta)\Omega}\,e^{\frac{3}{2}(2-\beta)F(\theta_iq^i)}
\end{equation*}
where $F(\theta_i q^i)=\theta_1 p_+ + \theta_2  p_- + \theta_3 p_\phi$, then we have

\begin{equation}
\frac{1}{3(2-\beta)}\left(
e^{3(2-\beta)\Omega}\right)^\bullet=e^{\frac{3}{2}(2-\beta)F(\theta_i
q^i)}\left[2p_0+\frac{144(2-\beta)}{\beta}\left(\frac{2^{\alpha-1}}{\alpha
} \right)^{\frac{1}{2\alpha-1}}\,p_\phi^{\beta}\left(t-t_{\alpha_i}
\right) \right]
\end{equation}
thus,

\begin{equation}
e^{3(2-\beta)\Omega}=a_0 +e^{\frac{3}{2}(2-\beta)F(\theta_i
q^i)}\,P_{\alpha,\beta}(t) \nonumber
\end{equation}
with $P_{\alpha,\beta}(t)=\left[6(2-\beta)p_0\left(t-t_{\alpha_i}
\right)+\frac{216(2-\beta)^2}{\beta}\left(\frac{2^{\alpha-1}}{\alpha
} \right)^{\frac{1}{2\alpha-1}}\,p_\phi^{\beta}\left(t-t_{\alpha_i}
\right)^2 \right]$, having

\begin{equation}
\Omega(t)=\frac{1}{3(2-\beta)}\ln \left| a_0
+e^{\frac{3}{2}(2-\beta)F(\theta_i q^i)}\,P_{\alpha,\beta}(t)
\right|, \label{sol-omega-nc}
\end{equation}
with $a_0$ an integration constant. Then, the volume of the universe
is
\begin{equation}
V(t)=e^{3\Omega}=\left( a_0 +e^{\frac{3}{2}(2-\beta)F(\theta_i
q^i)}P_{\alpha,\beta}(t) \right)^{\frac{1}{2-\beta}}. \label{volume}
\end{equation}
The anisotropic parameters $\beta_\pm(t)$ and scalar field $\phi(t)$
become

\begin{eqnarray}
\beta_\pm&=& {\beta_0}_\pm-2p_\pm \int \frac{dz}{a_0 + b_\alpha z +
c_\alpha z^2}+
\frac{36(2-\beta)}{\beta}\left(\frac{2^{\alpha-1}}{\alpha }
\right)^{\frac{1}{2\alpha-1}}\,p_\phi^{\beta}\left(t-t_{\alpha_i}
\right)\left\{\begin{tabular}{l}
$\theta_1, \quad for \,\beta_+$\\
$\theta_2, \quad for \,\beta_-$
\end{tabular}\right.,\nonumber\\
\phi(t) &=& \phi_{\alpha_i}+\left(\frac{2^{\alpha-1}}{\alpha }
\right)^{\frac{1}{2\alpha-1}}\,p_\phi^{\beta-1}\left[-24
+\theta_3\frac{36(2-\beta)}{\beta}p_\phi\right]\left(t-t_{\alpha_i}
\right),
\label{sol-phi-nc}
\end{eqnarray}
where the integrals are the same as those that appear in the
equation \eqref{parameter-ani}. As a test that these non-commutative
solutions are correct, when all $\theta_i=0$, we recover the
commutative solutions found in the previous section.

\section{Quantum commutativity}

The WDW equation for these models is obtained  by making the usual substitution
 $\Pi_{q^\mu}=-i \hbar \partial_{q^\mu}$ in \eqref{reducido} and promoting the classical Hamiltonian density in the
differential operator applied to the wave function $\Psi(\Omega,\beta_\pm,\phi)$, $\hat{\cal H}\Psi=0$; we have

\begin{equation}
\hat{\mathcal{H}}_c \Psi=\hbar^{2}
e^{-3(2-\beta)\Omega}\left[-\frac{\partial^2 \Psi}{\partial
\Omega^2} + Q\frac{\partial \Psi}{\partial \Omega}+ \frac{\partial^2
\Psi}{\partial \beta_+^2} +\frac{\partial^2 \Psi}{\partial
\beta_-^2}
   \right]
   -\frac{24}{\beta}\,\left(\frac{2^{\alpha-1}}{\alpha}
\right)^{\frac{1}{2\alpha-1}}\,\hbar^\beta \frac{\partial^\beta
\Psi}{\partial \phi^\beta}=0.
\label{reducido-wdw}
\end{equation}
a fractional differential equation of degree $\beta =
\frac{2\alpha}{2\alpha - 1}$, and employing the following ansatz for
the wave function $\Psi(\Omega,\beta_+,\beta_-,\phi) = \mathcal{
B}_{+}(\beta_+) \mathcal{B}_{-}(\beta_-)
\mathcal{A}(\Omega)\mathcal{F}(\phi)$ and next dividing between this
one, we have the following expressions:

\begin{eqnarray}
\frac{d^2 \mathcal{A}}{d\Omega^2}-Q\frac{d\mathcal{A}}{d\Omega}\pm
\left[ \ell_a^2 + \frac{\ell^2}{\hbar^2}e^{3(2-\beta)}\Omega
\right]\mathcal{A}&=& 0,
\label{wdw-omega}\\
\frac{d^\beta\mathcal{F}}{d \phi^\beta} \mp \frac{\ell^2 \beta}{24
\hbar^\beta}\left(\frac{\alpha}{2^{\alpha-1}}
\right)^{\frac{1}{2\alpha-1}} {\cal F} &=& 0, \label{wdw-phi}\\
\frac{d^2 \mathcal{B}_+}{d\beta_+^2} \pm \ell_+^2\mathcal{B}_+ &=& 0,
\label{wdw-beta+}\\
\frac{d^2 \mathcal{B}_-}{d\beta_-^2} \pm \ell_-^2{\cal B}_- &=& 0,
\label{wdw-beta-}
\end{eqnarray}
where $\ell^2$, $\ell_+^2$, $\ell_-^2$ are separations constants for the scalar field equation,
and the anisotropic function $\mathcal{B}_\pm$, respectively, also $\ell_a^2=\ell_+^2+\ell_-^2$.

Following the book of Polyanin~\cite{polyanin} (page 179.10), we
discovered the solution for the first equation for $\beta\not=2$,
considering different values in the factor ordering parameter and
both signs in the constant $\ell^2$, and the corresponding solutions
become

\begin{eqnarray}
\mathcal{A} & = & \mathcal{A}_0 e^{\frac{Q}{2}\Omega}\left\{
\begin{tabular}{ll}
$J_{\rho_-}\left[2\frac{\ell}{\hbar 3(2-\beta)}e^{\frac{3}{2}(2-\beta)\Omega} \right]$,&\quad for\, $+\ell^2$\\
$K_{\rho_+}\left[2\frac{\ell}{\hbar
3(2-\beta)}e^{\frac{3}{2}(2-\beta)\Omega} \right]$,&\quad for\,
$-\ell^2$
\end{tabular} \right. \label{sol-A} \\
\mathcal{B}_\pm &=& {\mathcal{B}_0}_\pm \left\{ \begin{tabular}{ll}
$\cos{\left(\ell_\pm \beta_\pm - \alpha_\pm \right)}$, &\qquad for \,
$+\ell_\pm^2$ \\
$\cosh{\left(\ell_\pm \beta_\pm + \alpha_\pm \right)}$, &\qquad for \,
$-\ell_\pm^2$ \end{tabular} \right. \label{sol-B}\\
\mathcal{F} &=& \left\{ \begin{tabular}{ll} $f(0)\mathcal{
E}_{\beta}\Big(\mp A\phi^\beta\Big)$, & for \,
$0<\beta \leq 1$ \nonumber\\
$f(0)\mathcal{E}_{\beta}\Big(\mp A\phi^\beta\Big) + f^\prime(0)
\sum_{n=0}^\infty \frac{(-1)^n A^n}{\Gamma(n\beta + 2)} \phi^{n\beta
+ 1}$, & for \,$1<\beta \leq 2$, \label{sol-F} \end{tabular}\right.
\end{eqnarray}
with the order in the Bessel function $\rho_\pm =\frac{\sqrt{Q^2
\pm4\ell_a^2}}{3(2-\beta)}$, $A = \mp \frac{\ell^2\beta}{24
\hbar^\beta}\Big(\frac{\alpha}{2^{\alpha -
1}}\Big)^{\frac{1}{2\alpha - 1}}$ and $\mathcal{E}_{\beta}$ is the
Mittag-Leffler function \cite{Haubold2011}.

Then, we have the probability density  $|\Psi|^2$ by considering
only $\beta \not= 2$, taking two sectors

\subsection{for $0<\beta \leq 1$}

\begin{figure}[ht]
\includegraphics[scale=0.35]{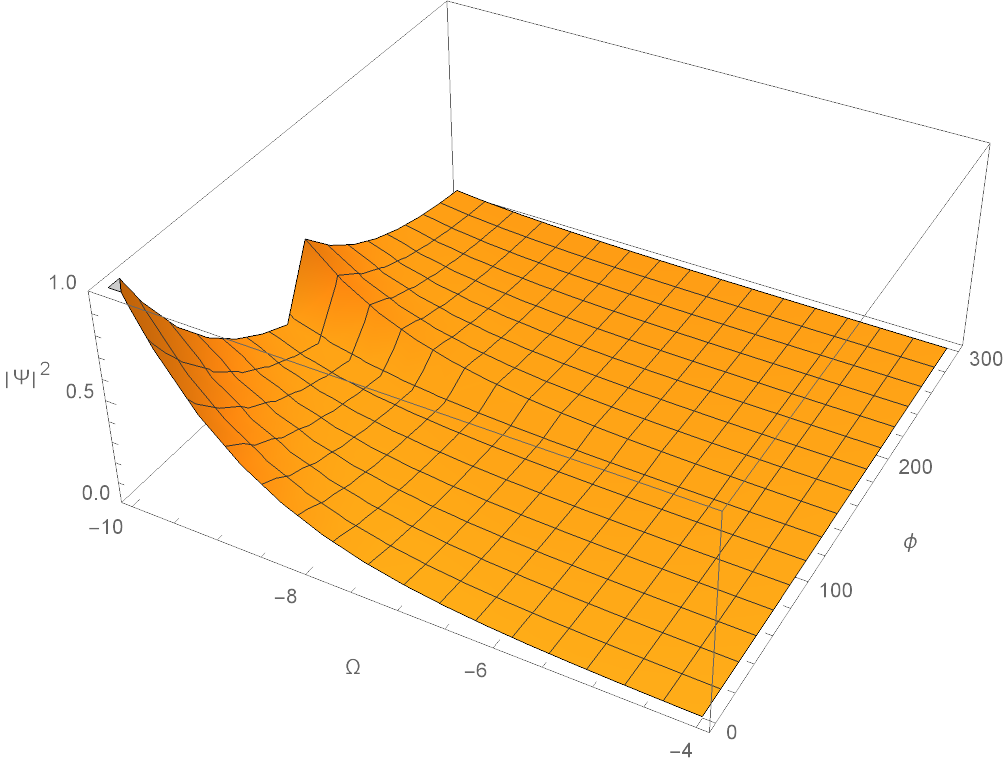}
\includegraphics[scale=0.35]{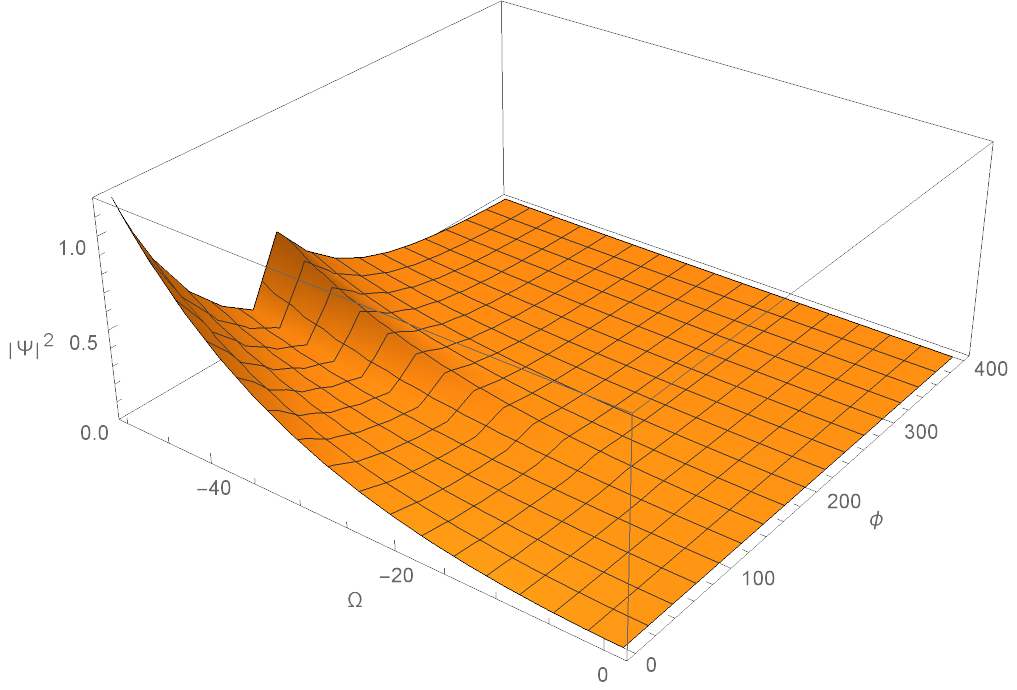}
\includegraphics[scale=0.35]{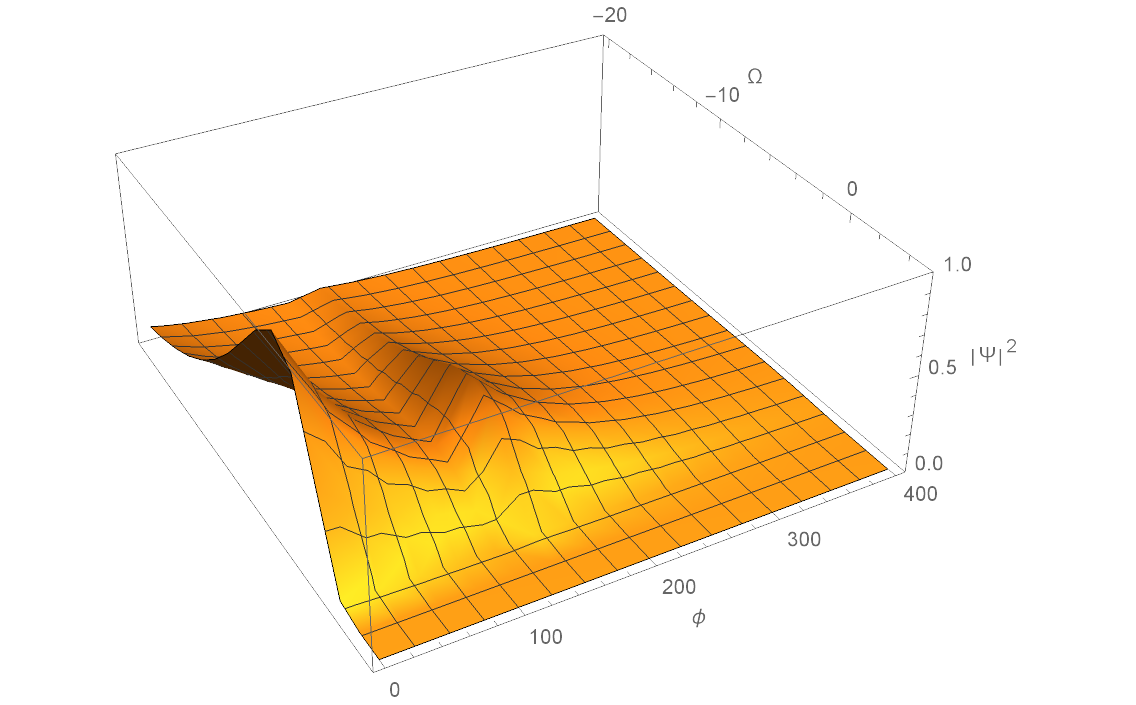}
\includegraphics[scale=0.35]{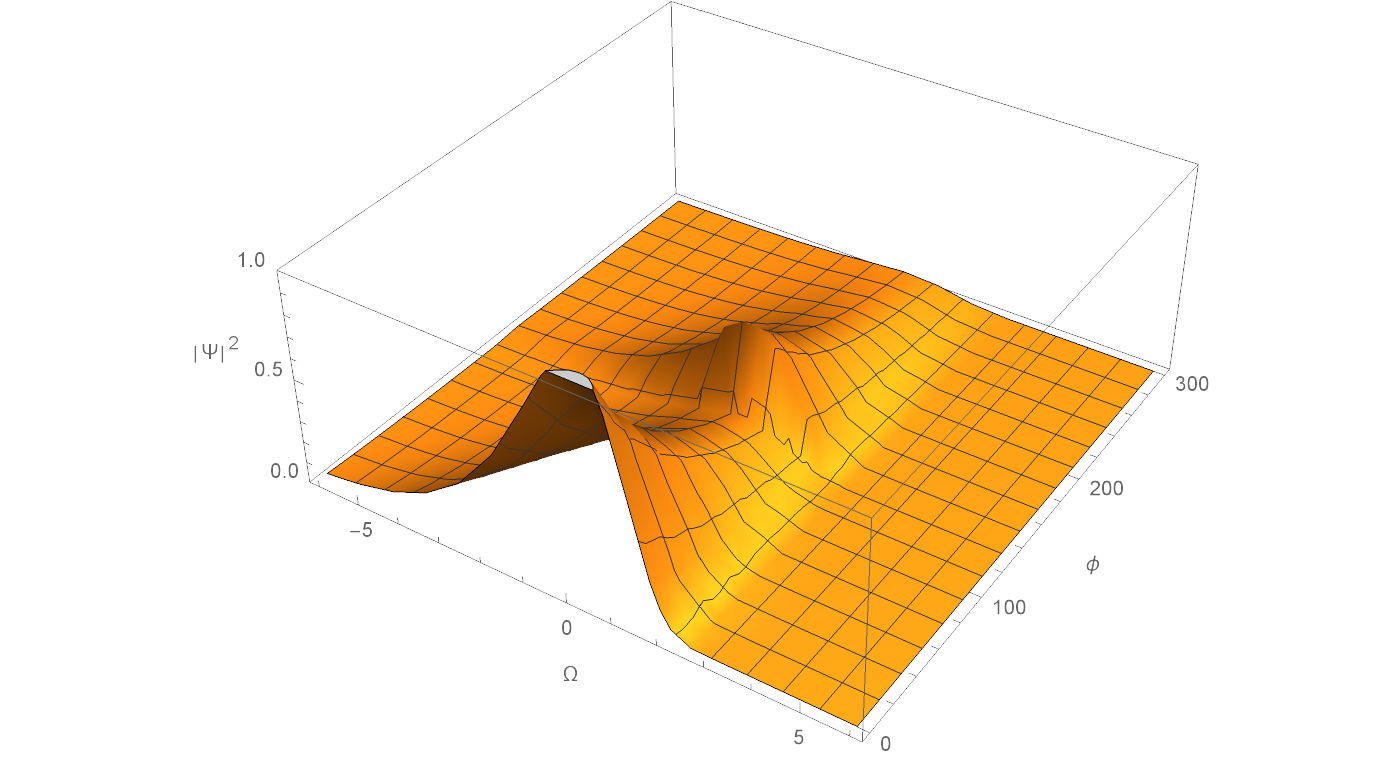}
\caption{%In the following plots, we
Shows the behaviour of the probability density, equation
\eqref{uno-c}, for dust era considering the modified Bessel
function. It is done by taking the values of the factor ordering
parameter $Q=-0.5, 0, 0.5, 2$, with the initial values
$\psi_0=\frac{\sqrt{5}}{500}, \frac{1} { 100}, \frac{\sqrt{5}}{5},
\frac{\sqrt{3}}{3}$, respectively. The values to the constants are;
$f(0)=f^\prime(0)=1$, $\ell=0.5$, $\ell_a=0.03$, and the order in
the modified Bessel function is $\rho_+=0.167862; 0.2; 0.167862;
0.666967,$ respectively.} \label{dust-conmutativo}
\end{figure}

In the figure \ref{dust-conmutativo}, we present the probability
density of the wave function (\ref{uno-c}) in the dust scenario
($\alpha \to \infty, \beta=1$), considering some values in the
factor ordering parameter $Q=-0.5, 0, 0.5, 2$, where we are taking
the initial values; $\psi_0=\frac{\sqrt{5}}{500}, \frac{1}{100},
\frac{\sqrt{5}}{5}, \frac{\sqrt{3}}{3}$, respectively. Employing the
modified Bessel function, which has decreasing behavior in the
direction of evolution of the scale factor like ($\Omega$),
presenting a structure in the direction of the scalar field, as two
parallel universes, where for the the value Q=2, this behavior is
more noticeable, making the scalar field relevant in quantum
evolution and remaining in the classical evolution, here the
anisotropic behavior is in the $\ell_a=0.03$ parameter, being very
small in these plots.

As the factor ordering parameter increases its values (small
values), the universe has a more uniform structure, just as the
normalization condition increases, presenting the non-normalized
behavior that characterizes quantum cosmology.

\begin{figure}[ht]
\includegraphics[scale=0.4]{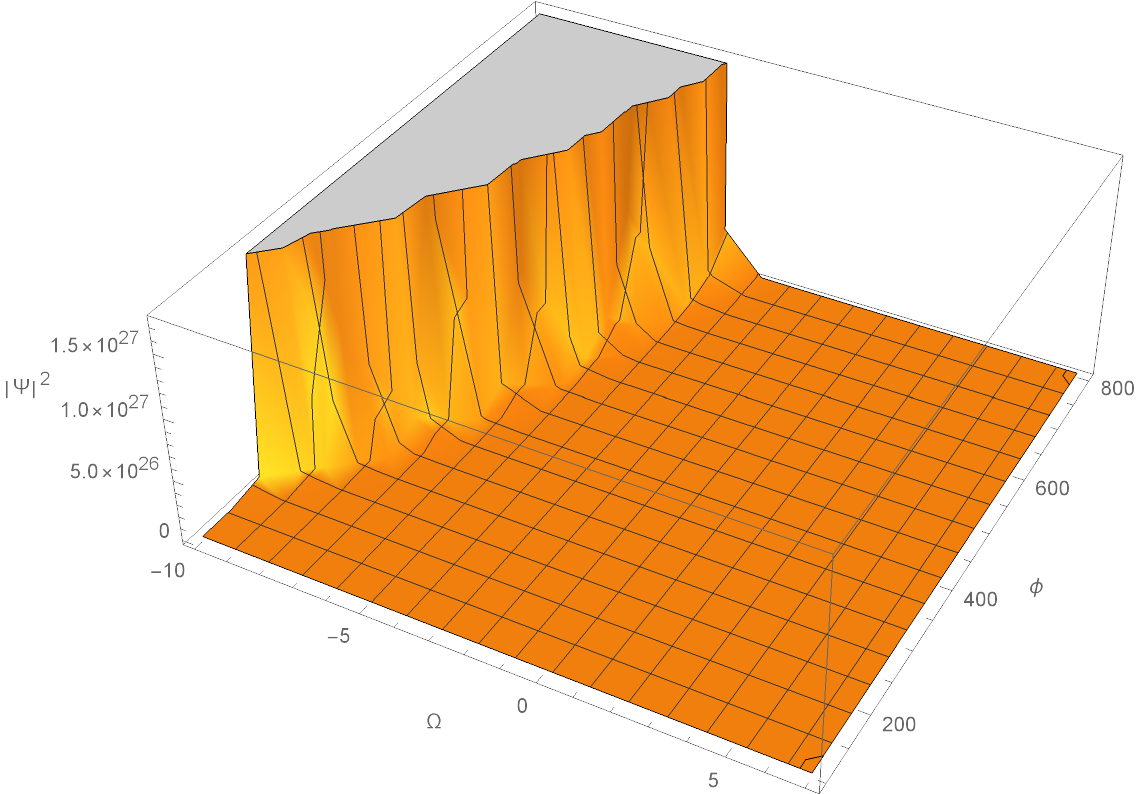}
\includegraphics[scale=0.4]{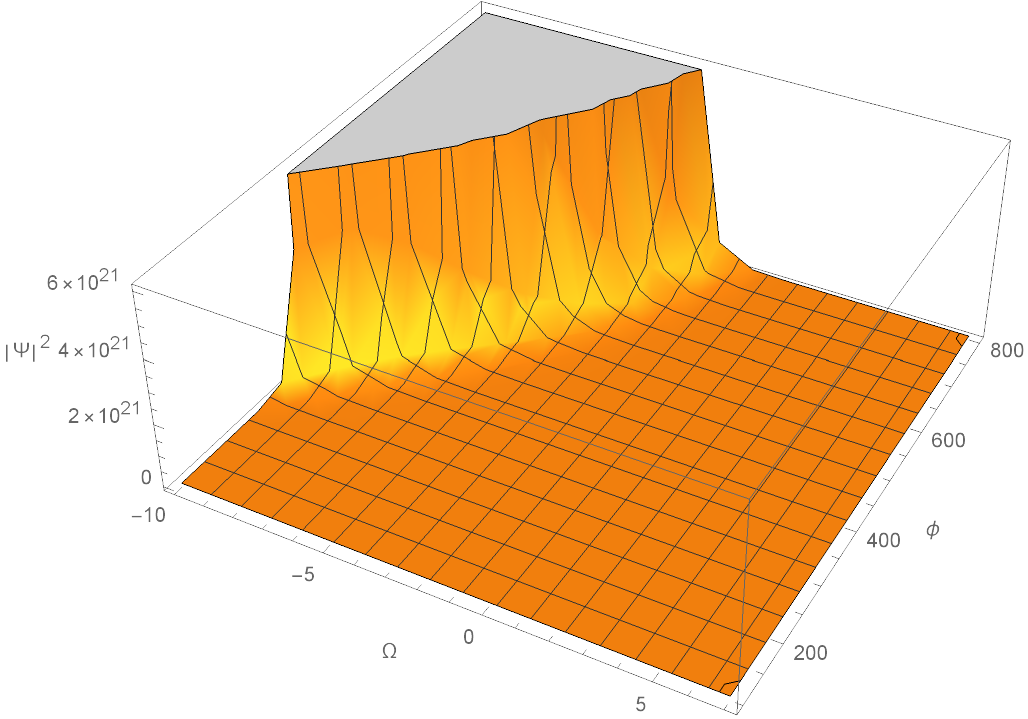}
\includegraphics[scale=0.4]{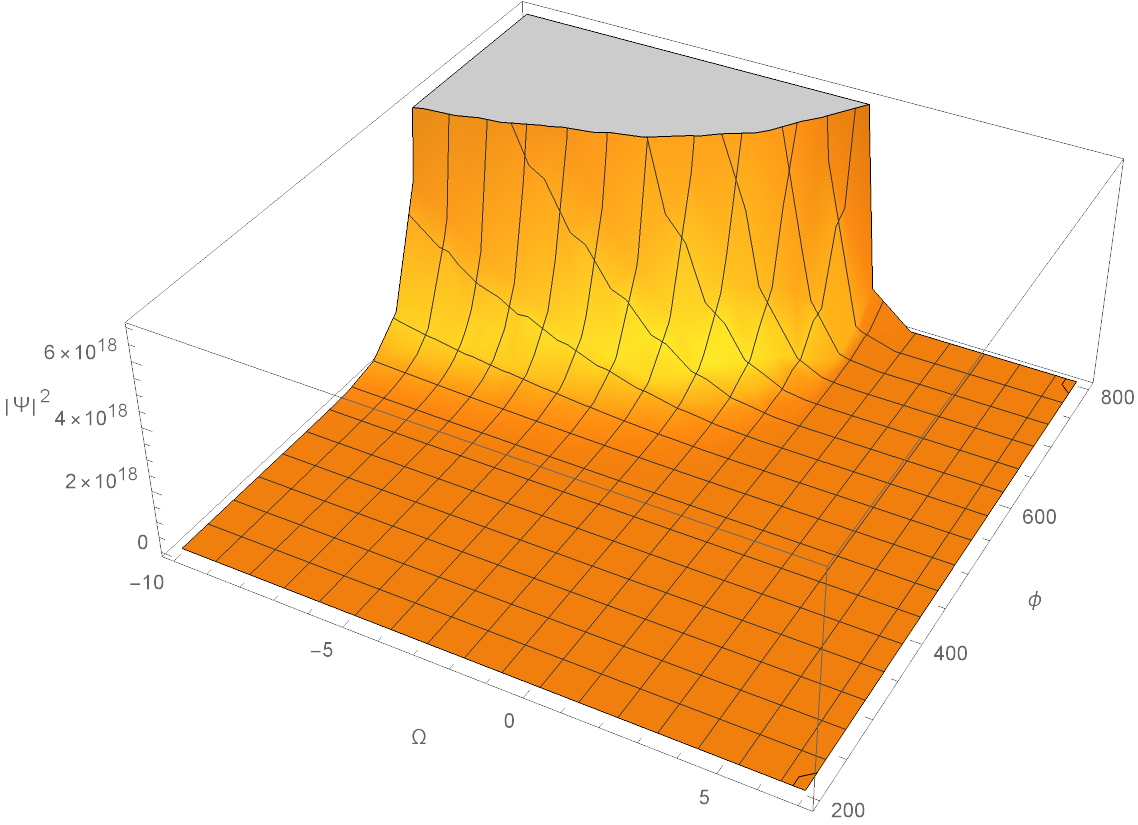}
\includegraphics[scale=0.4]{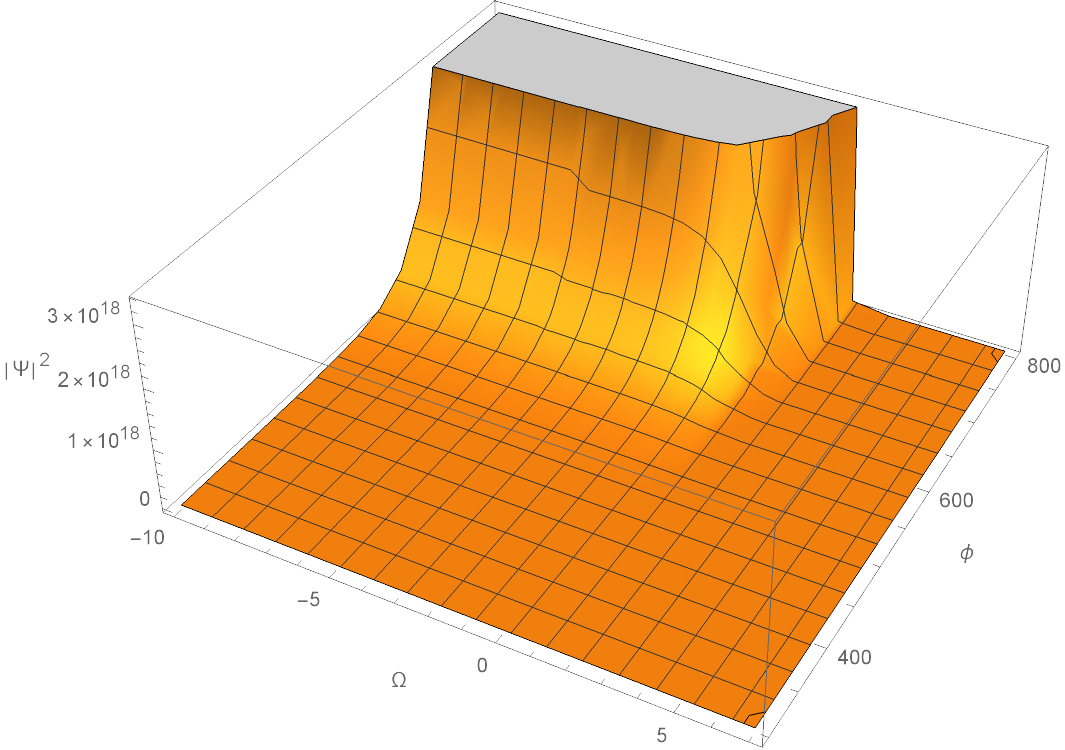}
\caption{Shows the behavior of the probability density, equation
\eqref{dos-c}, for radiation era, considering the modified Bessel
function. It is done by taking the values in the factor ordering
parameter $Q=-2,-1, 0, 2$, with the initial value
$\psi_0=\frac{\sqrt{5}}{500}, \frac{1} { 100}, \frac{\sqrt{5}}{5},
\frac{\sqrt{3}}{3}$, respectively. The values to the constants
$f(0)=f^\prime(0)=1$, $\ell=0.5$, $\ell_a=0.03$, and the order in
the modified Bessel function $\rho_+=1.00045,0.500899,0.03,1.00045$,
respectively.}\label{radiation-commutative}
\end{figure}

\begin{eqnarray}
|\Psi|^2&=&\psi_0^2\,e^{Q\Omega} f^2(0)\mathcal{
E}^2_{\beta}\Big(\mp A\phi^\beta\Big)
 \times
\nonumber\\
&& \times \left\{
\begin{tabular}{ll}
$\cos^2{\left(\ell_+ \beta_+ - \alpha_+ \right)}\cos^2{\left(\ell_-
\beta_- - \alpha_- \right)}
\,J^2_{\rho_-}\left[2\frac{\ell}{\hbar 3(2-\beta)}e^{\frac{3}{2}(2-\beta)\Omega} \right]$,& for\, $+\ell^2,+\ell_+^2,+\ell_-^2$\\
$\cosh^2{\left(\ell_+ \beta_+ + \alpha_+
\right)}\cosh^2{\left(\ell_- \beta_- + \alpha_-
\right)}K^2_{\rho_+}\left[2\frac{\ell}{\hbar
3(2-\beta)}e^{\frac{3}{2}(2-\beta)\Omega} \right]$,& for\,
$-\ell^2,-\ell_+^2,-\ell_-^2$
\end{tabular} \right. \label{uno-c}
\end{eqnarray}
 In the figure \ref{radiation-commutative}, we show the probability
density in the radiation stage of the universe, where it is
perceived that in the evolution of the universe the scalar field
enters late as the factor ordering parameter increases, which causes
this field to remain longer in the quantum world and, therefore,
remain as a background in the transition to the classical world,
just when the amplitude of the probability density decreases. This
behavior results in the dust stage, where this probability density
can be normalized manually (see figure \ref{dust-conmutativo}). Here
the behavior of the probability density of the wave function remains
in the non-normalized quantum cosmology format, which is why we do
not give a specific value to the initial value of $\psi_0$.

\subsection{for $1<\beta \leq 2$,}

\begin{eqnarray}
|\Psi|^2&=&\psi_0^2\,e^{Q\Omega}
\left[f(0)\mathcal{E}_{\beta}\Big(\mp A\phi^\beta\Big) + f^\prime(0)
\sum_{n=0}^\infty \frac{(-1)^n A^n}{\Gamma(n\beta + 2)} \phi^{n\beta
+ 1}\right]^2 \times
\nonumber\\
&& \times \left\{
\begin{tabular}{ll}
$\cos^2{\left(\ell_+ \beta_+ - \alpha_+ \right)}\cos^2{\left(\ell_- \beta_- - \alpha_- \right)}
\,J^2_{\rho_-}\left[2\frac{\ell}{\hbar 3(2-\beta)}e^{\frac{3}{2}(2-\beta)\Omega} \right]$,& for\, $+\ell^2,+\ell_+^2,+\ell_-^2$\\
$\cosh^2{\left(\ell_+ \beta_+ + \alpha_+
\right)}\cosh^2{\left(\ell_- \beta_- + \alpha_-
\right)}K^2_{\rho_+}\left[2\frac{\ell}{\hbar
3(2-\beta)}e^{\frac{3}{2}(2-\beta)\Omega} \right]$,& for\,
$-\ell^2,-\ell_+^2,-\ell_-^2$
\end{tabular} \right. \label{dos-c}
\end{eqnarray}

\section{Quantum non-commutativity}

The non-commutative version of fractional quantum cosmology presents
several difficulties due to the prefactor that remains associated
with the canonical moments that classically does not affect the
mathematical analysis, but when non-commutativity is used the
moments appear in that prefactor. The question that arises is: what
is the classical way of stating this prefactor that gives a better
 understanding of the quantum behavior of our problem? Since this prefactor can be separated into
 several parts and applied to a different sector of the Hamiltonian density. With this, the mathematical
 behavior is presented differently in each case. We will use the following separation of the classical
 non-commutative Hamiltonian density (\ref{hami-nc-n}) and proceed to perform the quantization using
 the correspondence principle.

\subsection{Separating the prefactor term in two parts}

The classical non commutative hamiltonian density
\begin{equation}
\mathcal{H}_{nc}= e^{-3(2-\beta)\left[\Omega -
\frac{\theta_1}{2}\Pi_+ -\frac{\theta_2}{2}\Pi_- -
\frac{\theta_3}{2}\Pi_\phi\right]}\left[\Pi_\Omega^2-
\Pi_+^2-\Pi_-^2
 \right] -
\frac{24}{\beta}\left(\frac{2^{\alpha-1}}{\alpha }
\right)^{\frac{1}{2\alpha-1}}\,\Pi_\phi^{\beta}, \label{hami-nueva}
\end{equation}
can be written as
\begin{equation}
\mathcal{H}_{nc}= e^{-3(2-\beta)\left[\Omega
\right]}\left[\Pi_\Omega^2- \Pi_+^2-\Pi_-^2
 \right] -
\frac{24}{\beta}\left(\frac{2^{\alpha-1}}{\alpha }
\right)^{\frac{1}{2\alpha-1}}\,e^{-3(2-\beta)\left[
\frac{\theta_1}{2}\Pi_+
 \frac{\theta_2}{2}\Pi_- +
\frac{\theta_3}{2}\Pi_\phi\right]}\Pi_\phi^{\beta}, \label{nueva}
\end{equation}
then, the corresponding quantum version, (including the factor
ordering problem), becomes
\begin{eqnarray}
&&\hbar^{2} e^{-3(2-\beta)\Omega}\left[-\frac{\partial^2
\Psi}{\partial \Omega^2} + Q\frac{\partial \Psi}{\partial \Omega}+
\frac{\partial^2 \Psi}{\partial \beta_+^2} +\frac{\partial^2
\Psi}{\partial \beta_-^2}
   \right]\nonumber\\
   &&-\frac{24}{\beta}\,\left(\frac{2^{\alpha-1}}{\alpha}
\right)^{\frac{1}{2\alpha-1}}\,\hbar^\beta
e^{3(2-\beta)\left[i\hbar\frac{\theta_1}{2}\frac{\partial}{\partial
\beta_+} + i\hbar\frac{\theta_2}{2}\frac{\partial}{\partial \beta_-}
+ i\hbar\frac{\theta_3}{2}\frac{\partial}{\partial
\phi}\right]}\frac{\partial^\beta \Psi}{\partial \phi^\beta}=0.
\label{q-wdw}
\end{eqnarray}
and employing the following ansatz for the wave function
$\Psi(\Omega,\beta_+,\beta_-,\phi)={\cal B_+}(\beta_+) {\cal
B_-}(\beta_-) {\cal A}(\Omega){\cal F}(\phi)$ and next dividing
between this one, we have the following expression
\begin{eqnarray}
&&\hbar^{2} e^{-3(2-\beta)\Omega}\left[\frac{-\frac{d^2 {\cal A}}{d
\Omega^2} + Q\frac{d {\cal A}}{d \Omega}}{{\cal A}} +\frac{
\frac{d^2 {\cal B}_+}{d \beta_+^2}}{{\cal B}_+} +\frac{ \frac{d^2
{\cal B}_-}{d \beta_-^2}}{{\cal B}_-}
   \right]\nonumber\\
   &&-\frac{24}{\beta}\,\left(\frac{2^{\alpha-1}}{\alpha}
\right)^{\frac{1}{2\alpha-1}}\,\hbar^\beta
\frac{e^{3(2-\beta)\left[i\hbar\frac{\theta_1}{2}\frac{d}{d
\beta_+}\right]}{\cal B}_+}{{\cal B}_+}
\frac{e^{3(2-\beta)\left[i\hbar\frac{\theta_2}{2}\frac{d}{d
\beta_-}\right]}{\cal B}_-}{{\cal B}_-} \frac{e^{3(2-\beta)\left[
i\hbar\frac{\theta_3}{2}\frac{d}{d \phi}\right]}\frac{d^\beta {\cal
F}}{d \phi^\beta}}{{\cal F}}=0. \label{wdw}
\end{eqnarray}
If we want to separate the differential equations, as a first
approximation we can consider that the terms that accompany the
differential equation of ${\cal F}$ are constrained to be constants,
nearly at unity, because the $\theta_i$ parameter is closed to zero
and subsequently use that information to solve the corresponding
differential equations as a system of equations.
\begin{equation}
\frac{e^{3(2-\beta)\left[i\hbar\frac{\theta_1}{2}\frac{d}{d
\beta_+}\right]}{\cal B}_+}{{\cal B}_+}=a_+^2 \approx 1, \quad
\frac{e^{3(2-\beta)\left[i\hbar\frac{\theta_2}{2}\frac{d}{d
\beta_-}\right]}{\cal B}_-}{{\cal B}_-}=a_-^2\approx 1,
\label{constra}
\end{equation}
(we choose for convenience $a_\pm^2$ and not $a_\pm$ in the last
equations) thus, equation (\ref{wdw}) is separated as
\begin{eqnarray}
&&\hbar^{2} e^{-3(2-\beta)\Omega}\left[\frac{-\frac{d^2 {\cal A}}{d
\Omega^2} + Q\frac{d {\cal A}}{d \Omega}}{{\cal A}} +\frac{
\frac{d^2 {\cal B}_+}{d \beta_+^2}}{{\cal B}_+} +\frac{ \frac{d^2
{\cal B}_-}{d \beta_-^2}}{{\cal B}_-}
   \right]\nonumber\\
   &&=\frac{24a_+^2 a_-^2}{\beta}\,\left(\frac{2^{\alpha-1}}{\alpha}
\right)^{\frac{1}{2\alpha-1}}\,\hbar^\beta
 \frac{e^{3(2-\beta)\left[
i\hbar\frac{\theta_3}{2}\frac{d}{d \phi}\right]}\frac{d^\beta {\cal
F}}{d \phi^\beta}}{{\cal F}} =\pm \ell^2=constant. \label{sep}
\end{eqnarray}
with this, we can obtain the following separated equations
\begin{eqnarray}
 \frac{d^2 {\cal B}_+}{d \beta_+^2} \pm \ell_+^2 {\cal
 B}_+ &=&0 ,\label{mas}\\
 \frac{d^2 {\cal B}_-}{d \beta_-^2} \pm \ell_-^2 {\cal
 B}_-&=& 0,\label{menos}\\
\frac{d^2 {\cal A}}{d \Omega^2} - Q\frac{d {\cal A}}{d \Omega}\pm
\left[\ell_a^2 + \frac{\ell^2}{\hbar^2}e^{3(2-\beta)\Omega}
\right]{\cal A}&=&0, \label{aa}\\
i\hbar \theta_3 \frac{3(2-\beta)}{2}\frac{d^{\beta+1} {\cal
F}}{d\phi^{\beta +1}} + \frac{d^\beta {\cal F}}{d\phi^\beta} \mp
\frac{\ell^2 \beta}{24\hbar^\beta a_+^2
a_-^2}\left(\frac{\alpha}{2^{\alpha -1}}
\right)^{\frac{1}{2\alpha-1}}{\cal F}&=&0. \label{phi-nc0}
\end{eqnarray}
The solution for the function ${\cal A}$ is the same that in the
commutative case, equation (\ref{sol-A}).

To solve the fractional equation (\ref{phi-nc0}), we write it as
follow:

\begin{equation}
\frac{d^{\beta + 1 }F}{d\phi^{\beta + 1}} + a\frac{d^\beta
F}{d\phi^\beta} + b F = 0 \label{phi1},
\end{equation}
where
\begin{eqnarray}
a &=&\frac{2 }{3i \hbar \theta_3(2 - \beta)}, \nonumber\\
b &=& \mp \frac{\ell^2 \beta }{24\hbar^\beta a_{+}^2
a_{-}^2}\Big(\frac{\alpha }{2^{\alpha - 1}} \Big)^{\frac{1}{2\alpha
- 1 }} \cdot \frac{2}{3i \hbar \theta_3(2 - \beta)}  = \pm
\frac{i\ell^2 \beta}{36  \hbar^{\beta + 1} \theta_3 (2 -
\beta)a_{+}^2 a_{-}^2  } \Big( \frac{\alpha}{2^{\alpha - 1}}
\Big)^{\frac{1}{2\alpha - 1}}. \nonumber
\end{eqnarray}
 Applying the Laplace transform to (\ref{phi1}), when $0<\beta \leq 2$, we have
\begin{equation}
s^{\beta + 1}F(s) - s^\beta f(0) - s^{\beta - 1} f^\prime(0) -
s^{\beta - 2} f^{\prime\prime}(0) + as^\beta F(s) - a s^{\beta -
1}f(0) - a s^{\beta - 2}f^\prime (0) + b F(s) = 0. \label{phi4}
\end{equation}
Solving with respect to $F(s)$, we obtain
\begin{eqnarray}
F(s) &=& f(0)\Big[\frac{s^\beta }{s^{\beta + 1} + as^\beta + b } +
\frac{as^{\beta - 1}}{s^{\beta + 1} + as^\beta + b}   \Big] +
f^\prime(0)\Big[\frac{s^{\beta-1} }{s^{\beta + 1} + as^\beta + b } +
\frac{as^{\beta - 2}}{s^{\beta + 1} + as^\beta + b}   \Big] +
\nonumber\\
&+& f^{\prime\prime}(0)\frac{s^{\beta - 2} }{s^{\beta + 1}+ as^\beta
+ b }. \label{phi5}
\end{eqnarray}

To take the inverse Laplace transform, we use the following formula
\cite{universe2024-nc}
\begin{equation}
\mathcal L^{-1}\Big[\frac{s^\gamma }{s^\alpha + as^\beta + b }
\Big] = t^{\alpha - \gamma - 1}\sum_{n=0}^\infty
\sum_{k=0}^\infty\frac{(-1)^{n+k} b^n a^k \binom{n+1+k}{k}
}{\Gamma[k(\alpha - \beta) + (n+1)\alpha - \gamma]} t^{k(\alpha -
\beta) + n\alpha} \label{formula}
\end{equation}
Taking the inverse Laplace transform in (\ref{phi5}), the solution
of the fractional differential equation (\ref{phi1}), for the case
$1<\beta \leq 2$, is given by
\begin{eqnarray}
{\cal F}_3(\phi) &=& f(0)\Big[\sum_{n=0}^\infty \sum_{k=0}^\infty \frac{(-1)^{n+k} b^n a^k  \binom{n+1+k}{k}}{\Gamma[(n+1)(\beta + 1) - \beta + k]} \Big] \phi^{n(\beta + 1) + k} +\nonumber\\
&+& a f(0) \Big[\sum_{n=0}^\infty \sum_{k=0}^\infty \frac{(-1)^{n+k} b^n a^k  \binom{n+1+k}{k}}{\Gamma[(n+1)(\beta + 1) - \beta + k + 1]} \Big] \phi^{n(\beta + 1) + k + 1} + \nonumber\\
&+& f^\prime(0) \Big[\sum_{n=0}^\infty \sum_{k=0}^\infty \frac{(-1)^{n+k} b^n a^k  \binom{n+1+k}{k}}{\Gamma[(n+1)(\beta + 1) - \beta + k + 1]} \Big] \phi^{n(\beta + 1) + k + 1} + \nonumber\\
&+& a f^\prime (0)  \Big[\sum_{n=0}^\infty \sum_{k=0}^\infty \frac{(-1)^{n+k} b^n a^k  \binom{n+1+k}{k}}{\Gamma[(n+1)(\beta + 1) - \beta + k + 2]} \Big] \phi^{n(\beta + 1) + k + 2}+ \nonumber\\
&+& f^{\prime\prime}(0) \Big[\sum_{n=0}^\infty \sum_{k=0}^\infty
\frac{(-1)^{n+k} b^n a^k \binom{n+1+k}{k}}{\Gamma[(n+1)(\beta + 1) -
\beta + k + 2]} \Big] \phi^{n(\beta + 1) + k + 2}. \label{phi6}
\end{eqnarray}
So, from here, we obtain the case, when $0<\beta \leq 1$,
\begin{eqnarray}
{\cal F}_1(\phi) &=& f(0)\Big[\sum_{n=0}^\infty \sum_{k=0}^\infty \frac{(-1)^{n+k} b^n a^k  \binom{n+1+k}{k}}{\Gamma[(n+1)(\beta + 1) - \beta + k]} \Big] \phi^{n(\beta + 1) + k} +\nonumber\\
&+& a f(0) \Big[\sum_{n=0}^\infty \sum_{k=0}^\infty \frac{(-1)^{n+k}
b^n a^k  \binom{n+1+k}{k}}{\Gamma[(n+1)(\beta + 1) - \beta + k + 1]}
\Big] \phi^{n(\beta + 1) + k + 1} +\nonumber \\ &+& f^{\prime}(0)
\Big[\sum_{n=0}^\infty \sum_{k=0}^\infty \frac{(-1)^{n+k} b^n a^k
\binom{n+1+k}{k}}{\Gamma[(n+1)(\beta + 1) - \beta + k + 1]} \Big]
\phi^{n(\beta + 1) + k + 1}. \label{phi7}
\end{eqnarray}
In the way in which the parameters of the differential equation for
the functional ${\cal F}$ are arranged, its solution for the $\theta
=0$ case does not reproduce the commutative case, since this
parameter remains in the denominator of the corresponding solution.
It is in this sense that non-commutativity makes the equations
involved in anisotropic fractional cosmology more complicated than
in the non-fractional case.

On the other hand, employing the constraints equations
(\ref{constra}) into equations (\ref{mas}) and (\ref{menos}), we
have
\begin{eqnarray}
 \frac{d^2 {\cal B}_+}{d \beta_+^2} \pm i\theta_1 \hbar\frac{3(2-\beta)\ell_+^2}{2a_+} \frac{d{\cal
 B}_+}{d\beta_+} \pm \frac{\ell_+2}{a_+^2}{\cal B}_+ &=&0 ,\label{mas-f}\\
 \frac{d^2 {\cal B}_-}{d \beta_-^2} \pm i\theta_2 \hbar\frac{3(2-\beta)\ell_-^2}{2a_-} \frac{d{\cal
 B}_-}{d\beta_-} \pm \frac{\ell_-2}{a_-^2}{\cal B}_-&=& 0,\label{menos-f}
\end{eqnarray}
the auxiliary polynomial for the first equation becomes
$$ r^2 \pm i\theta_1 \hbar\frac{3(2-\beta)\ell_+^2}{2a_+^2} r \pm
\frac{\ell_+2}{a_+^2}=0,$$ being the roots (in these roots the
quadratic terms in $\theta_j$ have been eliminated),
\begin{equation}
r_1= \mp i\theta_1 \hbar\frac{3(2-\beta)\ell_+^2}{4a_+^2} +
\sqrt{\mp \frac{ \ell^2_\pm}{a_+^2}},\qquad r_2= \mp i\theta_1
\hbar\frac{3(2-\beta)\ell_+^2}{4a_+^2} - \sqrt{\mp \frac{
\ell^2_\pm}{a_+^2}}.
\end{equation}

The corresponding solutions are:
\begin{equation}
\mathcal{B}_\pm = {\mathcal{B}_0}_\pm Exp\left[\mp i\theta_j \hbar
\frac{\ell_\pm^2}{a_\pm^2} \frac{3(2-\beta)}{4}\beta_\pm\right]\,\,
\left\{
\begin{tabular}{ll} $\cos{\left(\ell_\pm \beta_\pm - \alpha_\pm
\right)}$, &\qquad for \,
$+\ell_\pm^2$ \\
$\cosh{\left(\ell_\pm \beta_\pm + \alpha_\pm \right)}$, &\qquad for
\, $-\ell_\pm^2$ \end{tabular} \right. \label{sol-B-nc}
\end{equation}
and taking into account the limit of $\theta_j = 0$, this equation
is reduced to the commutative solutions (\ref{sol-B}).

Then, we have the probability density  $|\Psi|^2$ by considering
only $\beta \not= 2$, taking two sectors

\subsection{for $0<\beta \leq 1$}
\begin{figure}[ht]
\includegraphics[scale=0.4]{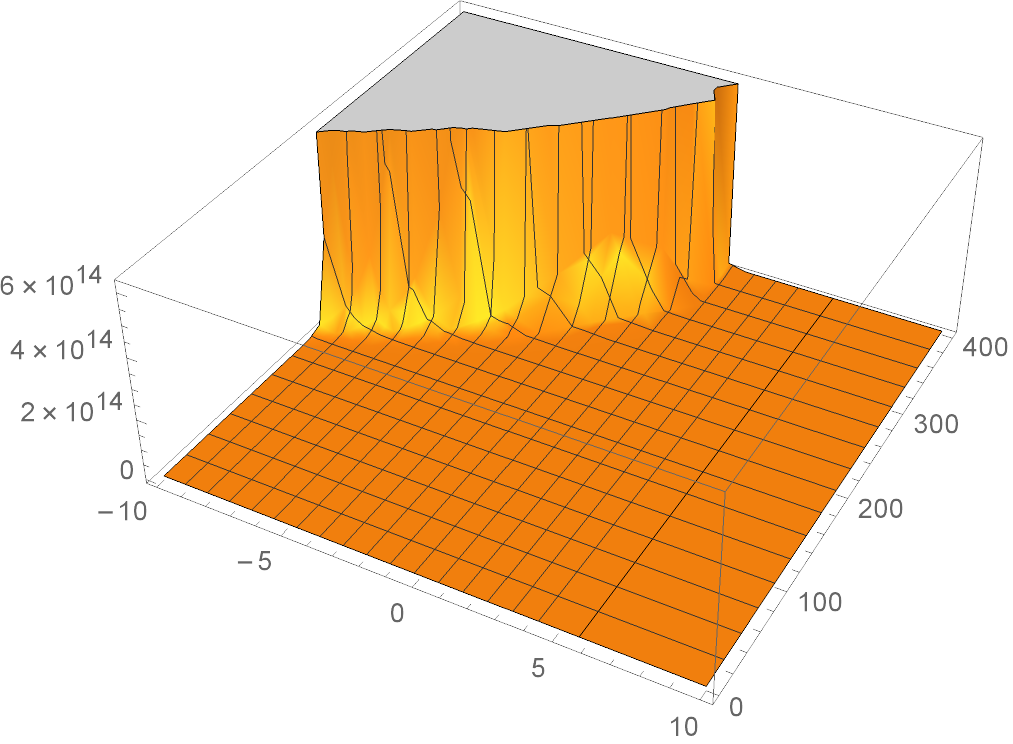}
\includegraphics[scale=0.4]{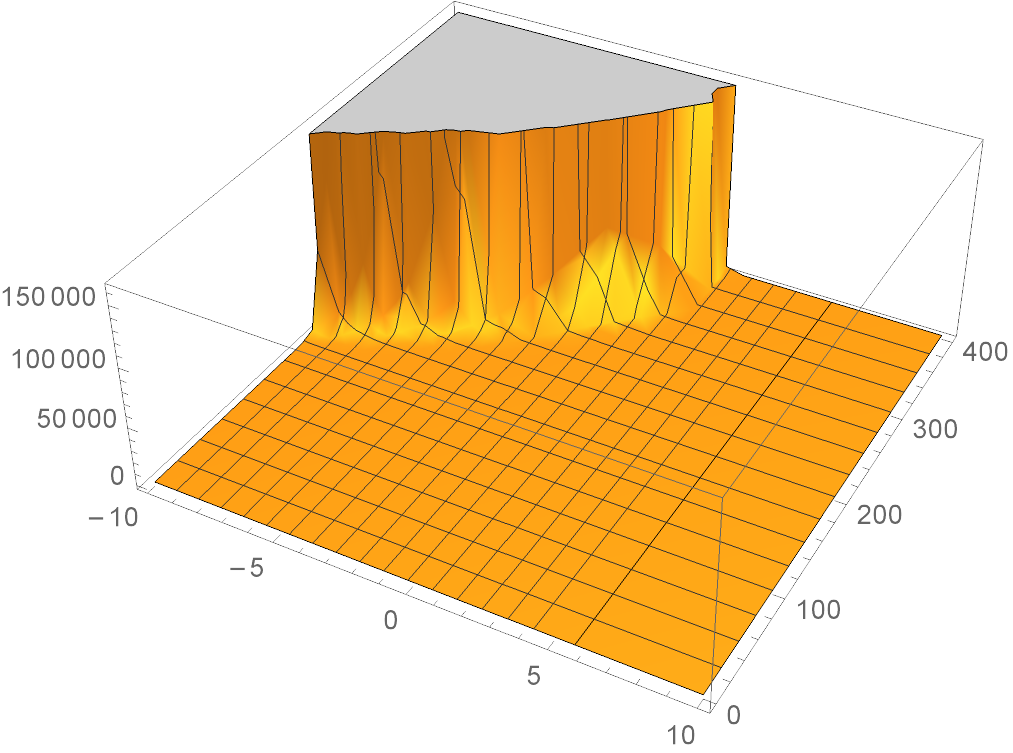}
\includegraphics[scale=0.4]{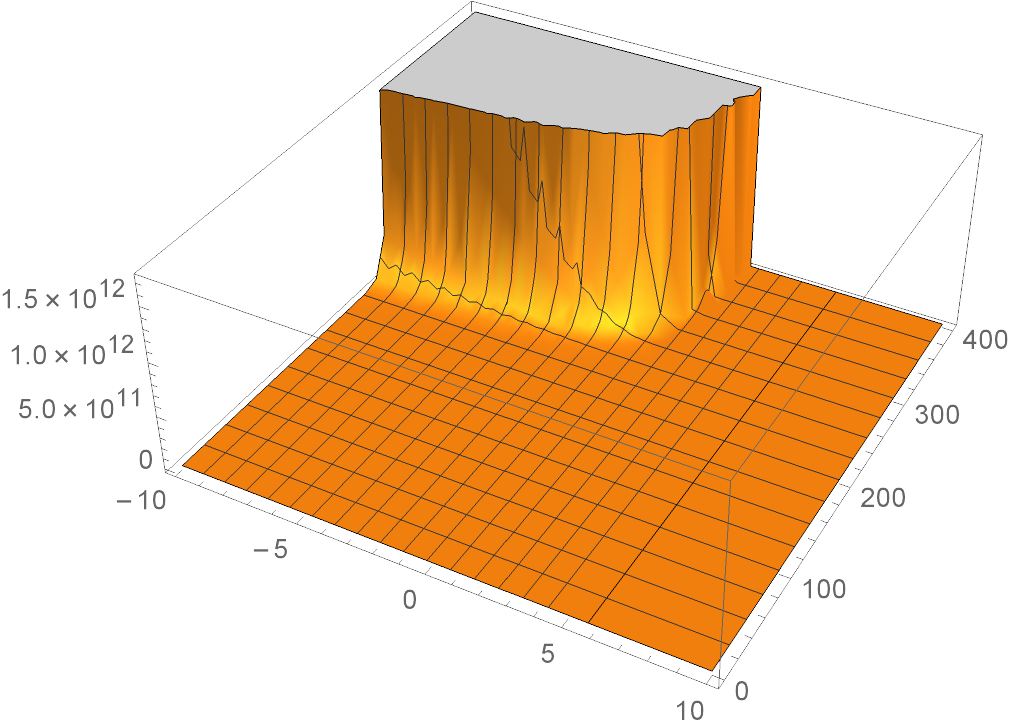}
\includegraphics[scale=0.4]{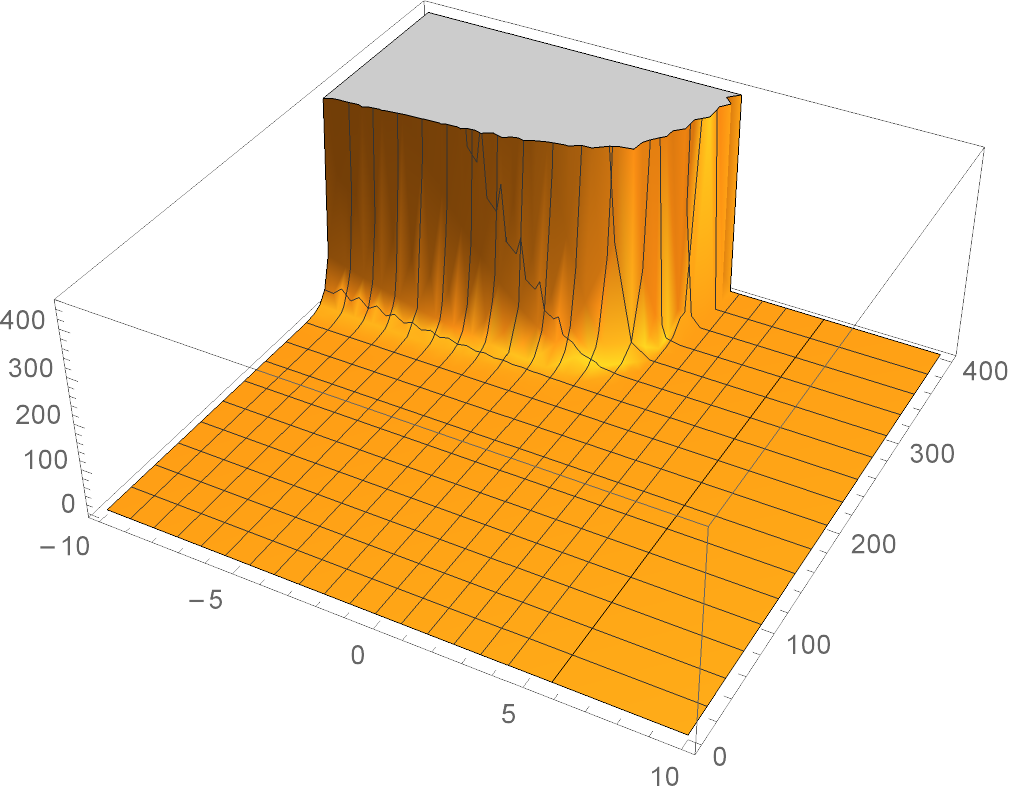}
\includegraphics[scale=0.4]{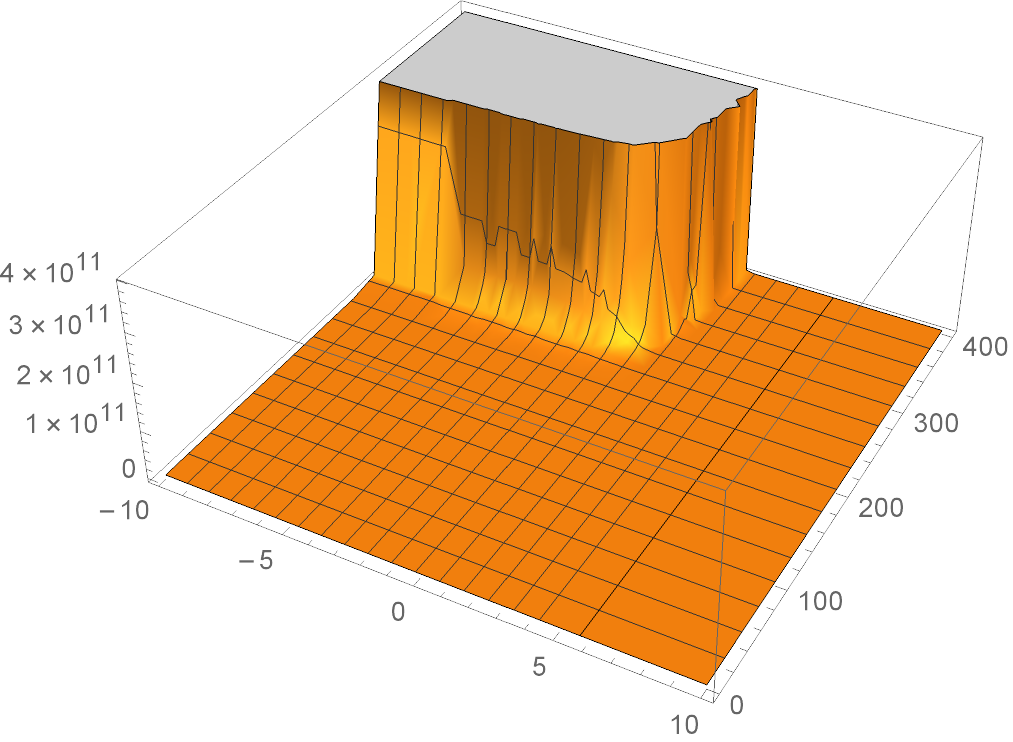}
\includegraphics[scale=0.4]{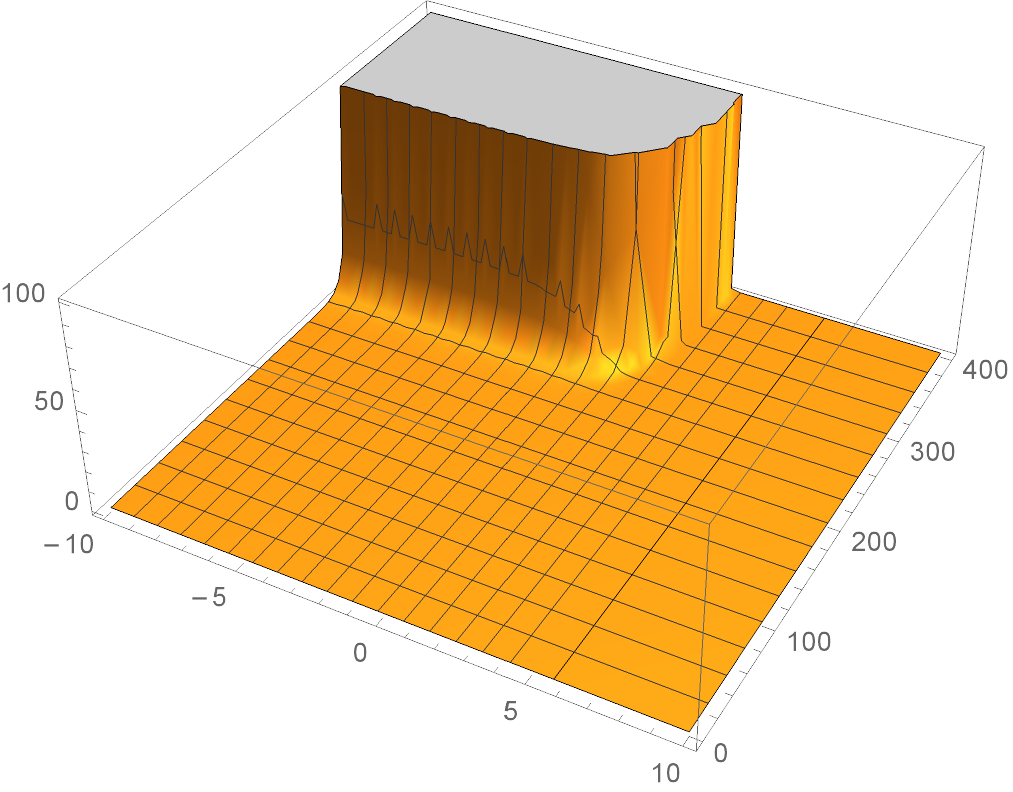}

\caption{In the following graphs, we show the behavior of the
probability density of the equation \eqref{uno-nc} for the dust era,
considering the modified Bessel function. Here, the values have been
taken in the factor ordering parameter $Q=-1,0,1$ and the
non-commutative parameter in the direction $\phi$, $\theta_3=0.1$
with $\psi_0=10^{- 26}$ and $\theta_3=0.9$ with $\psi_0=10^{-18}$.
We have taken the initial value and the constant values
$f(0)=f^\prime(0)=f^{\prime\prime}(0)=1$, $\ell=0.5$, $\ell_a =0.03
$, $a_+=0.0, a_-=0.95$, as well as, the order in the modified Bessel
function $\rho_+=0.333933,0.02$.}\label{nc-dust}
\end{figure}

\begin{figure}[ht]
\includegraphics[scale=0.4]{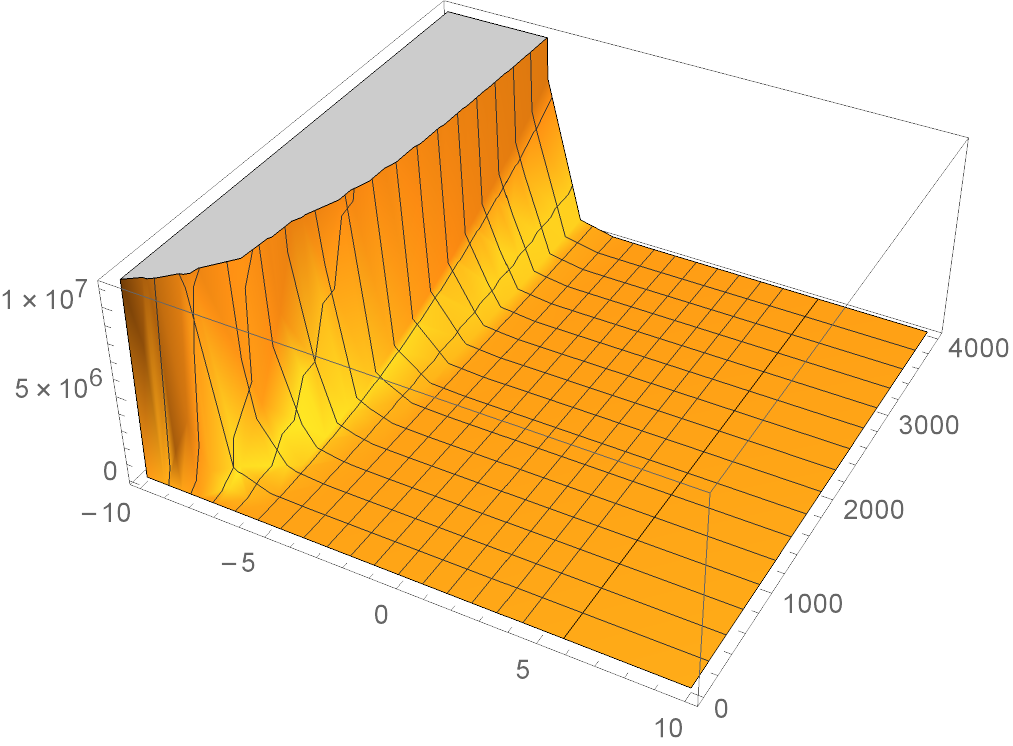}
\includegraphics[scale=0.4]{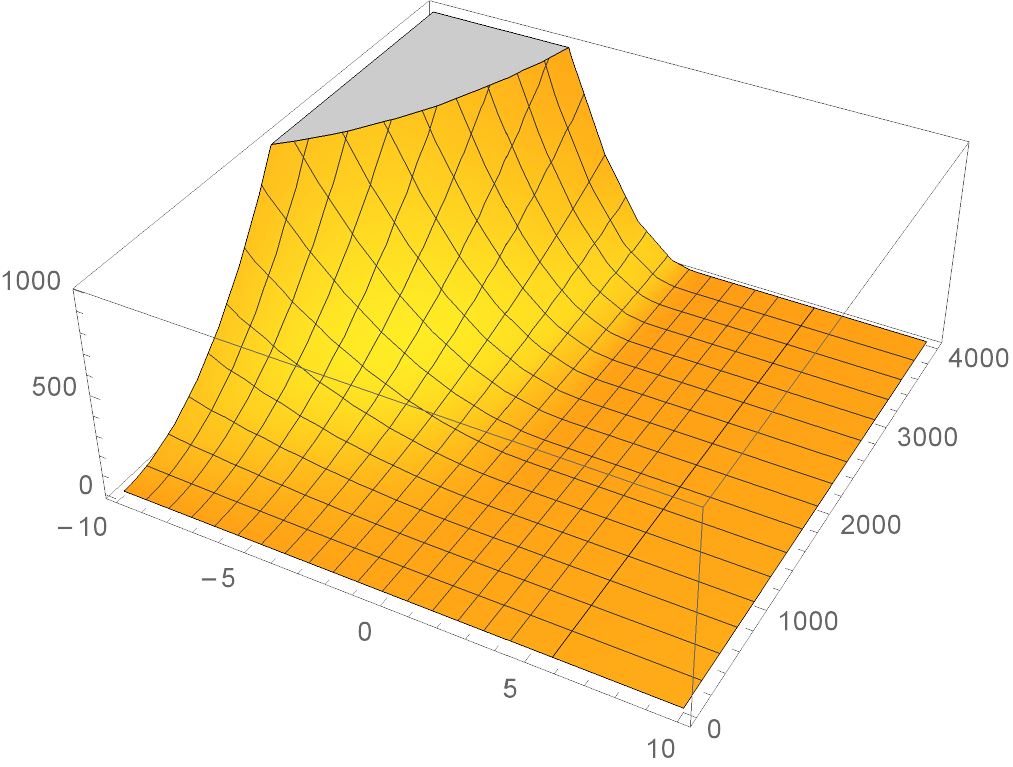}
\includegraphics[scale=0.4]{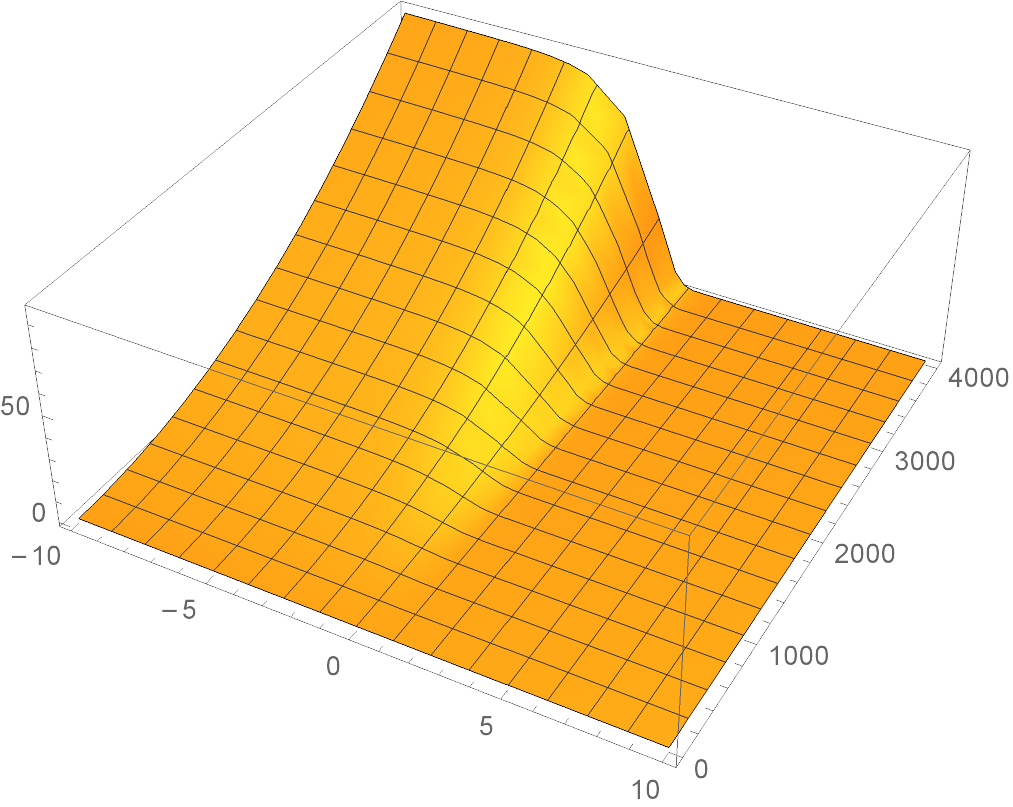}

\caption{In the following plots, we show the behavior of the
probability density, equation \eqref{uno-nc}, for dust era,
considering the modified Bessel function. Here, the values have been
taken in the factor ordering parameter $Q=-1,0,1$ and the non
commutative parameter in the direction with $\phi$ and
$\theta_3=9\times 10^9$. We have taken the initial value and the
constant values $f(0)=f^\prime(0)=f^{\prime\prime}(0)=1$,
$\ell=0.5$, $\ell_a=0.03$, $a_+=0.9, a_-=0.95$, $\psi_0=10^{-3}$ and
the order in the modified Bessel function
$\rho_+=0.333933,0.02$.}\label{nc-dust-infinity}
\end{figure}

\begin{eqnarray}
|\Psi|^2&=&\psi_0^2\,e^{Q\Omega}\, {\cal F}_1^2\times
\nonumber\\
 && \times
\left\{
\begin{tabular}{ll}
$e^{-\Theta_i}\,\cos^2{\left(\ell_+ \beta_+ - \alpha_+
\right)}\cos^2{\left(\ell_- \beta_- - \alpha_- \right)}
\,J^2_{\rho_-}\left[2\frac{\ell}{\hbar 3(2-\beta)}e^{\frac{3}{2}(2-\beta)\Omega} \right]$,& for\, $+\ell^2$\\
$e^{\Theta_i}\,  \,\cosh^2{\left(\ell_+ \beta_+ + \alpha_+
\right)}\cosh^2{\left(\ell_- \beta_- + \alpha_-
\right)}K^2_{\rho_+}\left[2\frac{\ell}{\hbar
3(2-\beta)}e^{\frac{3}{2}(2-\beta)\Omega} \right]$,& for\, $-\ell^2$
\end{tabular} \right. \label{uno-nc}
\end{eqnarray}
with $e^{\pm \Theta_i}=Exp\left[\pm
i\hbar\frac{3(2-\beta)}{2}\left[\theta_1 \frac{\ell_+^2}{a_+^2}
\beta_++\theta_2 \frac{\ell_-^2}{a_-^2} \beta_-\right]\right]$

\subsection{for $1<\beta \leq 2$}
\begin{figure}[ht]
\includegraphics[scale=0.4]{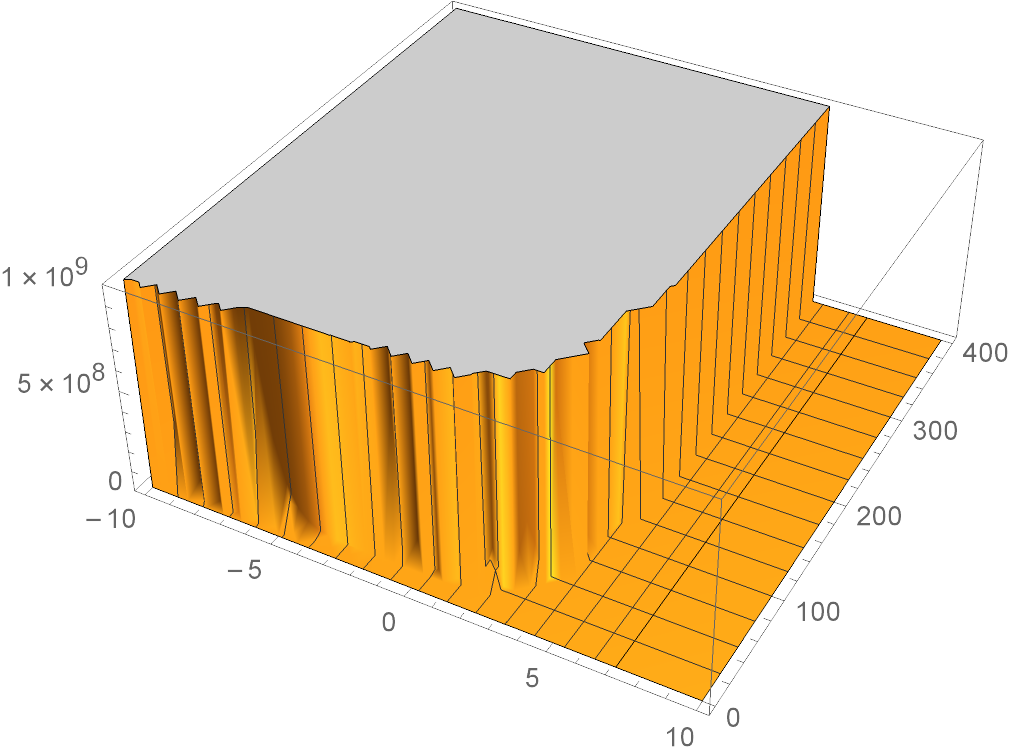}
\includegraphics[scale=0.4]{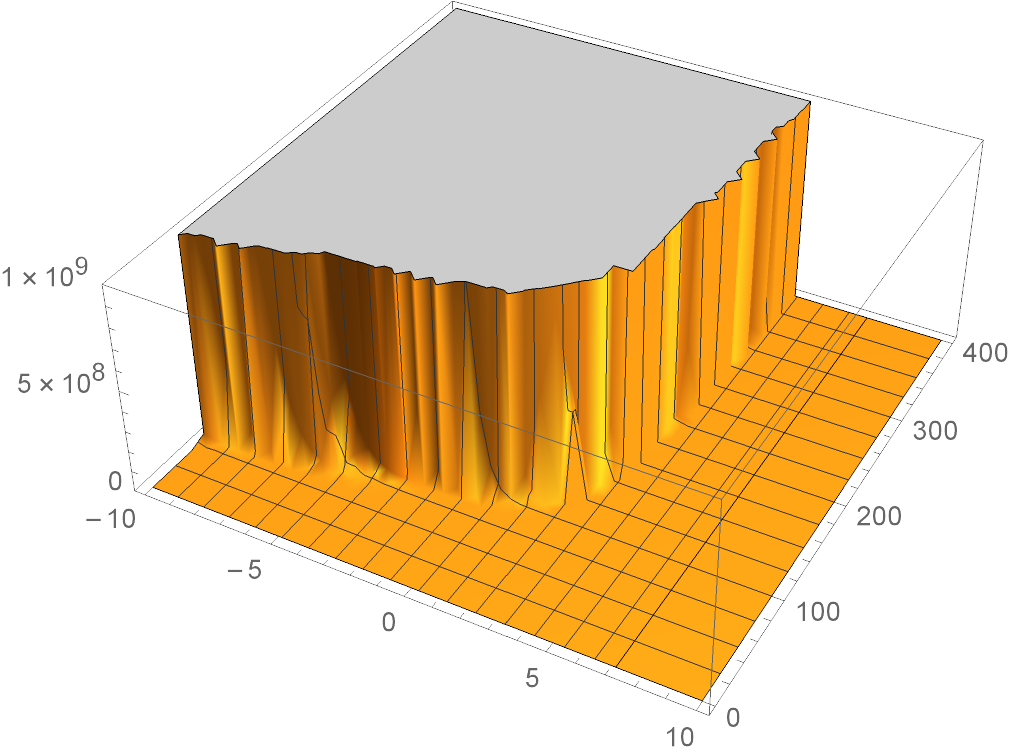}
\includegraphics[scale=0.4]{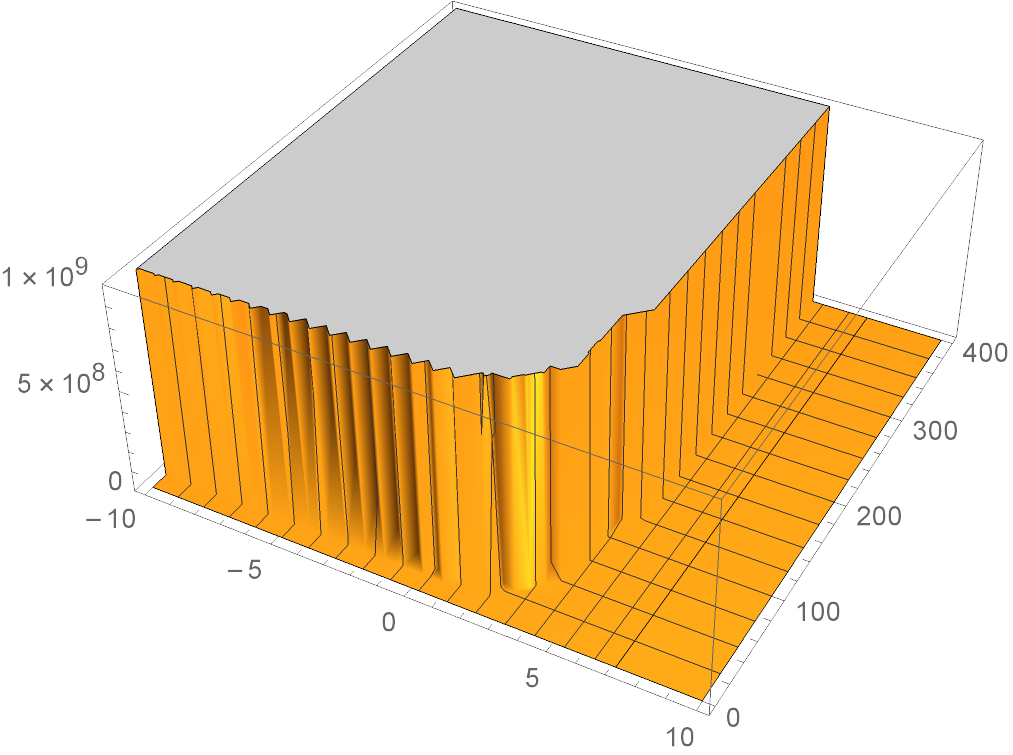}
\includegraphics[scale=0.4]{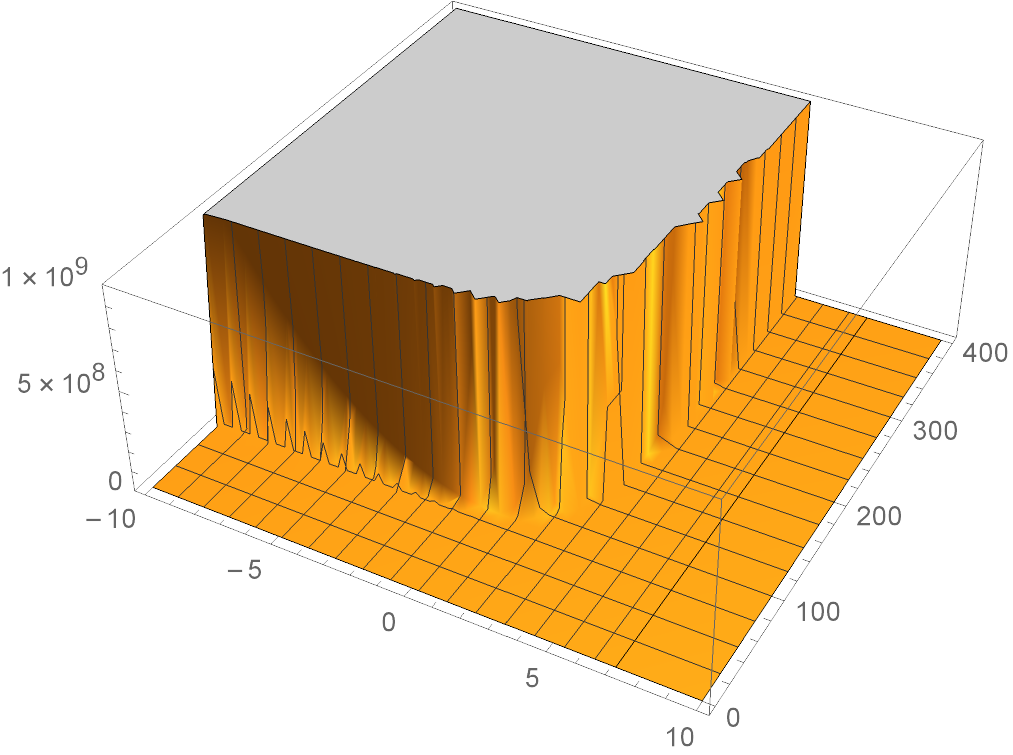}
\includegraphics[scale=0.4]{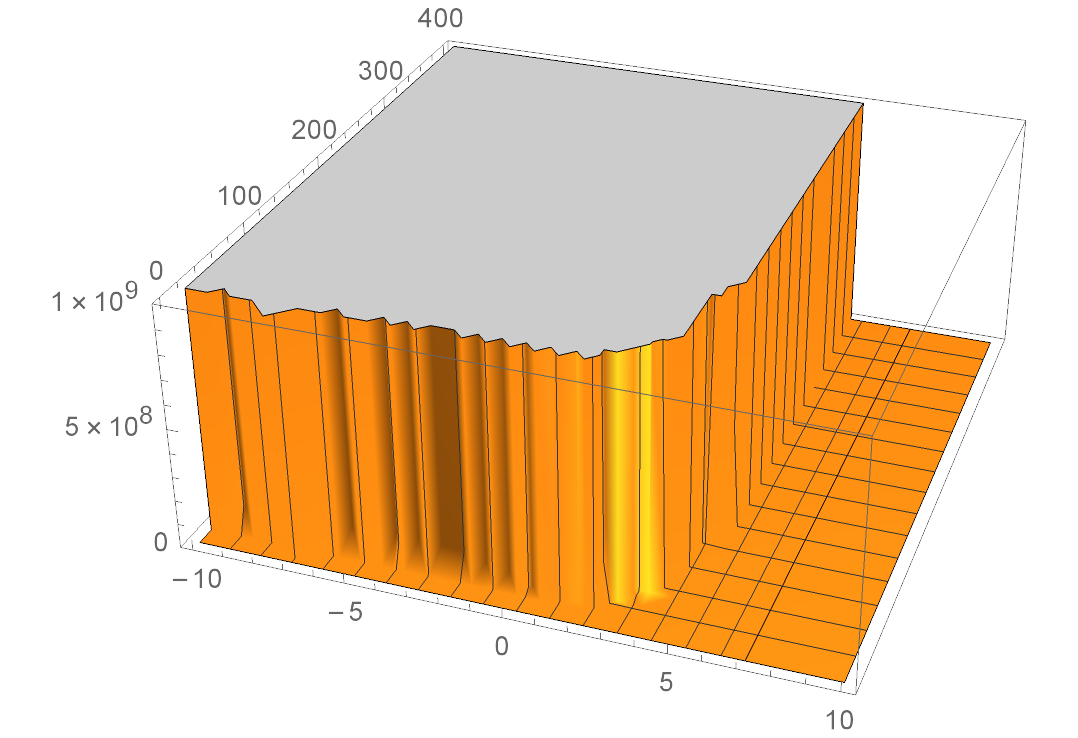}
\includegraphics[scale=0.4]{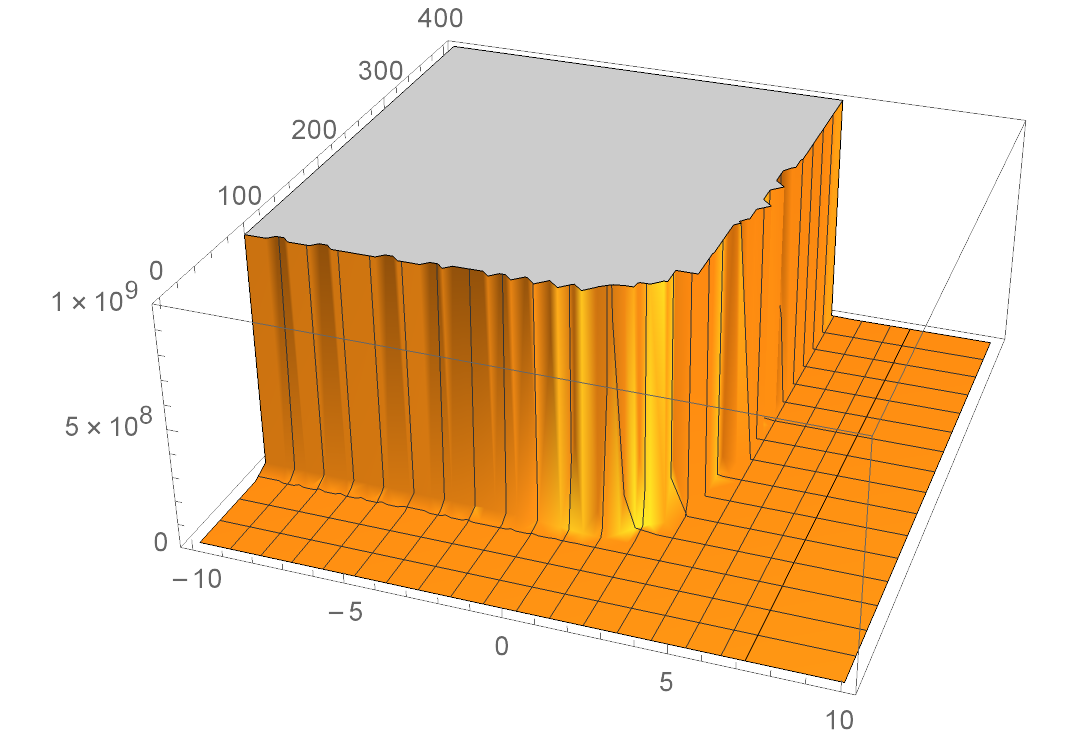}

\caption{In the following plots, we show the behavior of the
probability density, equation\eqref{dos-nc}, for radiation era,
considering the modified Bessel function. It is done by taking the
values of the factor ordering parameter $Q=-1,0,1$, and the non
commutative parameter in the direction $\phi$, $\theta_3=0.1, 0.9$.
We have taken the initial value and the constant values
$f(0)=f^\prime(0)=f^{\prime\prime}(0)=0.1$, $\ell=0.5$,
$\ell_a=0.03$, $a_+=0.0, a_-=0.95$ and the order in the modified
Bessel function $\rho_+=0.500899,0.03$.}\label{nc-radiation}
\end{figure}

 In figure \ref{nc-radiation}, we show the probability density in
the radiation stage of the non commutative universe, where it is
perceived that the evolution of the universe, the global effect of
the non-commutativity between the field coordinates of the system in
the fractional quantum cosmology scheme causes the probability
density to shift or shrink in the opposite direction in both fields
$(\Omega, \phi)$, causing the classical universe to emerge sooner,
which would mean that the current universe should have more time
than is usually mentioned, as in mentioned in the reference
\cite{Jalalzadeh2023}, employing the fractional framework. Here the
behavior of the probability density of the wave function remains in
the format of quantum cosmology, not normalized, which is why we do
not give a specific value to the initial value of $\psi_0$.

\begin{eqnarray}
|\Psi|^2&=&\psi_0^2\,e^{Q\Omega}\, {\cal F}_3^2 \times
\nonumber\\
&& \times \left\{
\begin{tabular}{ll}
$e^{-\Theta_i}\,\cos^2{\left(\ell_+ \beta_+ - \alpha_+
\right)}\cos^2{\left(\ell_- \beta_- - \alpha_- \right)}
\,J^2_{\rho_-}\left[2\frac{\ell}{\hbar 3(2-\beta)}e^{\frac{3}{2}(2-\beta)\Omega} \right]$,& for\, $+\ell^2$\\
$e^{\Theta_i}\,\,\cosh^2{\left(\ell_+ \beta_+ + \alpha_+
\right)}\cosh^2{\left(\ell_- \beta_- + \alpha_-
\right)}K^2_{\rho_+}\left[2\frac{\ell}{\hbar
3(2-\beta)}e^{\frac{3}{2}(2-\beta)\Omega} \right]$,& for\, $-\ell^2$
\end{tabular} \right. \label{dos-nc}
\end{eqnarray}

Other ways of separating this prefactor do not lead us to anything
plausible, since when we make $\theta \to 0$, the values of the
parameters on which the roots of a cubic equation depend diverge,
then the function ${\cal B}_\pm$ is indeterminate, so we decided not
to put those cases.

\section{Conclusions}
From the quantum perspective of this work, some theories applied to
fractional cosmology in the K-essence formalism become very
complicated, because a prefactor survives that does not appear in
non-fractional cosmology, since it is absorbed in the choice of the
appropriate gauge. Commutative behaviors cannot be reproduced from
these solutions, since the non-commutative parameter enters the
evolution of the functions in the denominator, and can diverge in
this limit. We only report the cases that we consider most plausible
from a physical point of view.

In this sense, the combination of fractional differential equations
and non-commutativity in anisotropic models seems to be not a good
combination, since when solving the resulting quantum equations,
this combination causes the domain of definition of the
non-commutative parameter to change to another domain of definition
that does not allow us to viably recover the commutative quantum
world, as are shown in the figures 3,4 and 5. In this way,  we
discarded the other choices of the prefactor given that due to the
way in which the non-commutative parameter is worked, it seems that
we are facing an ambiguity or separation of definition space.

For the very particular case treated in this work, the quantum
evolution of the probability density in the radiation stage is very
similar to the standard evolution, but with the novelty that the
scalar field enters later in that evolution according to the value
of factor ordering parameter increases, and the amplitude of this
density decreases, which means that in the dust era of the universe
this evolution gives us the idea of two parallel universes, a
product of the fractional process. The fact that the scalar field
enters later means that it remains as a classical background for
longer in the evolution of the universe, possibly with large values,
between stage and stage of the universe.

In the classical fractional sector, the non-commutative parameter is
appearing in the scalar field relations, since that is where the
fractional behavior comes from, and not from the scale factor
solutions (up to an exponential prefactor of non-commutative
parameters), as it usually appears in non-fractional commutative
cosmology, making The behavior of these functions is different in
both formalisms. However here, we can recover the classical
commutative world from the non-commutative sector.

\acknowledgments{ J.S. was partially supported by PROMEP grants
UGTO-CA-3 and CIIC-118/2024. All authors were partially supported by
SNI-CONACyT. J.J.R appreciates the support of DAIP, within the
framework of the project {\it Sobre la Dimensionalidad de los
Operadores Diferenciales Fraccionarios}: grant 203/2024;   We thank
Itzayana A.M.R for illuminating this work. This work is part of the
collaboration within the Instituto Avanzado de Cosmolog\'ia. Many
calculations were done by Symbolic Program REDUCE 3.8.}

\end{document}